\documentclass{aa}  

\usepackage{graphicx}
\usepackage[export]{adjustbox}
\usepackage[dvipsnames]{xcolor}
\usepackage{txfonts}
\bibpunct{(}{)}{;}{a}{}{,}
\newcommand{\LA}[1]{#1}
\begin{document}

    \title{Effects of accretion on the Structure and Rotation of Forming Stars}
   \subtitle{}

   \author{L. Amard
          \inst{1,2}
          \and
          S. P. Matt\inst{2}
          }

   \institute{AIM, CEA, CNRS, Universit\'e Paris-Saclay, Universit\'e de Paris, Sorbonne Paris Cit\'e, F-91191 Gif-sur-Yvette, France
            \and
   University of Exeter, Physics and Astrophysics dept, Stoker Road, EX44QL, Exeter, UK\\
            \email{louis.amard@cea.fr}
             }

   \date{Submitted, 1st Review}

 
  \abstract
   {Rotation period measurements of low-mass stars show that the spin distributions in young clusters do not exhibit the spin-up expected due to contraction, during the phase when a large fraction of stars are still surrounded by accretion discs. Many physical models have been developed to explain this feature based only on different type of star-disc interactions. During this phase the stars accrete mass and angular momentum and may experience accretion enhanced-magnetised winds. The stellar structure and angular momentum content thus strongly depend on the properties of the accretion mechanism. At the same time, the accretion of mass and energy has a significant impact on the evolution of stellar structure and moment of inertia. Thus our understanding of young-stars spin rates requires a description on how accretion affects stellar structure and angular momentum simultaneously.}
   {In this study we aim to understand the role of accretion in explaining the observed rotation rate distributions of forming stars.}
   {We compute evolution models of accreting very young stars and determine, in a self-consistent way, the effect of accretion on stellar structure and the angular momentum exchanges between the stars and their disc. 
   We then vary the deuterium content, the accretion history, the entropy content of the accreted material, and the magnetic field as well as the efficiency of the accretion-enhanced winds.}
   {The models are driven alternatively both by the evolution of the momentum of inertia, and by the star-disc interaction torques. Of all the parameters we tested, the magnetic field strength, the accretion history and the Deuterium content have the largest impact. The injection of heat only plays a major role early in the evolution.}
   {This work demonstrates the importance of the moment of inertia's evolution under the influence of accretion to explain the constant rotation rates distributions that are observed over the star-disc interactions. When accounting for rotation, the models computed with an up-to-date torque along with a consistent structural evolution of the accreting star are able to explain the almost constant spin evolution for the whole range of parameter we investigated, albeit only reproducing a narrow range around the median of the observed spin rate distributions. Further development including for example more realistic accretion histories based on dedicated disc simulations are likely needed to reproduce the extremes of the spin rate distributions.}

   \keywords{Stars:rotation --
                stars:pre-main sequence --
                stars:low-mass -- 
                stars:formation --
                accretion, accretion discs
               }

   \maketitle
%

\section{Introduction}

Understanding the formation stage of low-mass stellar system is of crucial importance for both planetary and stellar science. 
The early pre-main sequence (PMS) is the stage for many important events in a stellar system and is still poorly understood despite a lot of efforts on the topic in the last few decades.
The total angular momentum of the system is particularly important because it is a main driver of the early stellar evolution and the formation of the proto-planetary disc. 

During the first stages of its evolution, classical stellar evolution theory predicts that the star contracts while it relaxes to a stable state, and thus should spin up. In addition, the accretion of material coming from the inner edge of the proto-planetary disc should increase the stellar angular momentum. Following these basic assumption, the presence of a disc around a forming star should be, in theory, associated with a very fast stellar rotation rate.
However, observations of stars in clusters younger than $\sim$10 Myr show that the distributions of rotation-rates are approximately constant in time, or at least are inconsistent with the systematic spinning-up that is expected.  
Furthermore, stars that are currently surrounded by accretion discs do not systematically exhibit faster rotation than those without.
These apparent paradoxes are often described using a pseudo-explanation called ``disk locking,'' referring to a highly simplified concept that the presence of a disc keeps a star ``locked'' to a constant rotation rate during its evolution \citep{Koenigl91,Edwards93,Bouvier93,Rebull2004,Bouvier2007,KennedyKenyon2009,BouvierPPVI,Davies2014,Venuti17,Rebull2018,Rebull2020,Rebull2022}.
To describe the observed behavior, several theoretical explanations have been proposed to extract angular momentum from the systems. We can mention the X-winds model \citep{Shu1994,FPA2000}, the accretion-powered stellar winds \citep[APSW]{MattPudritz2005,MattPudritz2008}, or the magnetospheric ejections \citep{ZanniFerreira2009,ZanniFerreira2013}. While almost none of these mechanisms are exclusive by essence they have rarely been studied together \citep{Gallet2019}. In addition, the systems are very dynamic and complicated that a lot of variable parameters must be considered simultaneously and ideally self-consistently. 
Theoretical models for angular momentum transfer between the star and its environment (accretion disks and/or winds) show that the net torque on the star can be either negative or positive, and that the value depends on stellar and environmental properties that are themselves evolving with time.  
Therefore, an understanding of the evolution of stellar angular momentum requires a description of several properties evolving together.

The modelling of low-mass rotational evolution has been steadily and consistently improved over the last decade \citep{GB13,GB15,LS15,Amard2016,Somers2016,Amard2020,SL20}.
However very little work has been dedicated to the very early evolution, when the star still has a disc and seem to keep an almost constant rotation rate. In fact, all the above-mentioned models assume a constant rotation rate for a given duration which is associated to the time of interaction with the surrounding proto-planetary disc.
A few groups studied the effect of variable duration of this Star-Disc Interaction (SDI) phase to reproduce the rotation periods on longer timescale \citep{VB15,Coker2016,Roquette2021}.
We can also mention the work by \citep{SiessLivio97} on the role of a reduced convective mixing to explain the very slow rotation in young clusters, and the work by \cite{BatyginAdams13} on the link between rotation rates during the SDI phase and misalignment of close-in Jupiter mass planets.
Finally, a few rotation evolution model have accounted for a physical evolution during the SDI phase but require extreme stellar properties to explain the distributions \citep{Matt2010,Matt2012a,Gallet2019} or remain empirical \citep{Johnstone2014}.

Because of the stellar contraction, the typical rotational evolution of young stars is strongly influenced by the evolution of the moment of inertia. In order to account for its variation, rotation evolution models make use of stellar evolution tracks to follow the evolution of the stellar structure. The models start the evolution around 1 Myr or after the disc-coupling phase.  Thus the early structure evolution is often neglected. 
The community has been using models which do not account for any accretion process by either \cite{Baraffe98} or by \cite{Siess2000}.
The radius and thus moment of inertia can be drastically affected by accretion of material with different composition and thermodynamic properties \citep{Siess1997,Baraffe2009,Hosokawa2011,Kunitomo2017,Kunitomo2018}. 

Another element SDI has been shown to play a role on is the amount of lithium observed in young clusters and metal-poor old stars. It was shown by \cite{Bouvier2018} to already be connected to the rotation rates of PMS stars before the first five million years. \cite{Eggenbergeretal12} showed that a longer disc-coupling time generates a larger shear at the base of the convective zone which would then efficiently transport the lithium to deeper part of the star where it can be burnt.
On lithium evolution still, \cite{Baraffe2017} showed that the entropy content of the accreted material affects the temperature at the base of the convective region, and thus the amount of Lithium burnt during the PMS. We can also mention \cite{Tognelli2020,Tognelli2021} who tested the impact of several accretion scenario on the abundance of lithium and more generally of light elements.

In addition, accretion leaves a mark on the stellar structure that can remain for several tens of Myr. Its impact on the stellar pulsation spectra of $\delta$Scuti stars was estimated by \cite{Steindl2021,Steindl2022,Steindl2022Nature}.
On longer timescale, \cite{Zhang2019} solved the long standing dilemma on the solar measured abundances and the current stellar structure estimated by helioseismology. They showed that changing the composition of the accreted material during the first Myr was able to change the internal composition of the Sun and satisfy both the spectroscopic and the helioseismic constrains.

The early pre-main sequence is a highly dynamical phase and is therefore hard to model on evolutionary timescales. A common approach is to start from short time span multidimensional MHD simulations \citep{MattPudritz2005,MattPudritz2008,Romanova13,Romanova18,Romanova21,ZanniFerreira2009,ZanniFerreira2013,Pantolmos2020,Ireland2021,Ireland2022}, and to derive scaling laws to use in 1D evolution models on longer timescale.
In this work, we will be using the results from the star-disc PLUTO simulations by \cite{Ireland2022} and implement them in the STAREVOL evolution code. 

The model and the implementation are described in section 2. Section 3 presents some non-rotating accreting models to validate the evolution compared to other recent work. In section 4, we discuss the rotational evolution and test various parameters of the torque laws that we finally compare to the rotation rate distribution in open clusters. Finally, in section 5, we recall the limitations of the model, draw conclusions and discuss the forthcoming possibilities.

\section{Description of the model}
    \subsection{Stellar evolution model : STAREVOL with accretion}
        \subsubsection{Standard case}
We use the 1D stellar evolution code STAREVOL \citep{Siess2000,PalaciosCharbonnel2006,Lagarde2012,Amard2019} and refer to these papers for the specific details of the code.
Our models are computed in 1D single stars without magnetic field but under the influence of rotation and accretion starting from a small initial seed. STAREVOL solves the structure equations and assume hydrostatic equilibrium at each timestep assuming that the effect of rotation on the structure are treated as a perturbation \LA{on the hydrostatic and radiative equilibrium equations} \citep[\LA{for more details, see}][]{ES76,MeynetMaeder1997}. This assumption is valid as long as we remain in a relatively slow rotating regime with respect to the break-up velocity. \LA{Along this work, all the models we consider remain below half of the break-up velocity and mostly around 10 to 20\%. This leads to an increase of radius of a few percent at most.}
We patch atmospheric $T(\tau)$ relations from \cite{KrishnaSwamy} from $\tau=0.005$ to $\tau=100$. We kept this set of models for numerical stability reasons but verified in advance that the results were qualitatively not too different with PHOENIX atmosphere models as in \cite{Amard2019}.
The temperature gradient is computed following the mixing-length theory \citep[MLT,][]{CoxGiuli68} with a fixed parameter $\alpha_{\rm MLT}$ = 2.173 that reproduces the solar luminosity and radius at 4.57 Gyr. The internal chemical composition changes with the nuclear reactions and the diffusion in the convective envelope given by the MLT. We do not account for any overshooting nor any microscopic diffusion.
Although STAREVOL is capable of computing differentially rotating models solving an the full set of angular momentum equation \citep{Zahn1992,MathisZahn2004,Palacios2003,DecressinMathis2009}, as a first step we keep the star as a solid body both in the radiative and the convective regions. The effect of differential rotation will be investigated in further work. \LA{Since the star is in a state of solid body rotation, there are no chemical transport by shear mixing, and the transport associated to the meridional circulation is negligible in the current case. Similarly we do not account for any transport by magnetic instabilities nor by waves.} 
While we consider a magnetic field to compute the spin evolution, the effects of magnetism on the structure \citep{SP2015} or the internal transport of angular momentum \citep{Eggenberger2022} are neglected in the current case. 

The initial composition is set as given in \cite{Asplund2021}, except for the fraction of deuterium which is variable.
As was emphasised by \cite{Kunitomo2017}, the Deuterium is the main nuclear driver of the evolution during the early pre-main sequence, in particular in the case of low-entropy accretion. However, its initial value is still relatively uncertain and the literature varies between 10ppm and 40ppm \citep{Hebrard2005,Linsky2006}. Because of how quickly the deuterium is consumed, we account for diffusion when solving the associated nuclear reactions to determine the deuterium profile.

        \subsubsection{Accretion}
        \label{Sect:FormAcc}
Stellar evolution may drastically change when accreting material.
Accretion modifies the stellar structure by bringing mass, energy and new chemicals to the star. Some energy is kept by the star (as thermal energy), and some is directly radiated away.
The question of the deposition of the accreted material in the star is very important for the evolution, we refer to \cite{SiessForestini96,Siess1997,Hartmann1998,Baraffe2009,Kunitomo2017} for detailed discussions on the physics of accretion in stellar evolution models. In this work we follow the simpler description by \cite{Hartmann1998} that we will now recall.

We parametrise the heat injected by the accreted material into the forming star as
\begin{equation}
L_{\rm add} = \xi \frac{GM_\star\dot{M}_{\rm acc}}{R_\star} 
\end{equation}
with $M_\star$ and $R_\star$ are the mass and the radius of the central star, respectively, $\dot{M}_{\rm acc}$ is the accretion rate and $\xi$ is the accretion efficiency such that $0 \leq \xi \leq 0.5$. We set the upper limit at 0.5 and not 1 to account for the radiative cooling from the disc surface and the geometry of the problem. The star is only accreting on a small surface, generally at mid latitude.

An additional term comes in the Energy equation to account for the accreted material.
Here we follow again \cite{Kunitomo2018} and the energy is only deposited in the upper layer of the star with a fractional mass of $m_{ke}$.  We then have the expression of the energy deposited per unit mass 
\begin{equation}
    \varepsilon_{\rm add} = \frac{L_{\rm add}}{M_\star}\times {\rm Max}\left[0;\frac{2}{m_{ke}^2}\left(\frac{M_r}{M_\star}-1 + m_{ke}\right)\right],
\end{equation}
with $M_r$ the mass coordinate and $m_{ke}$ is a fraction of the total mass. In this work we assume $m_{ke}=0.02$. Although much more simplistic, this somewhat mimics the energy deposition in the upper layer given by the more physical formalism by \cite{SiessForestini96} and confirmed by the 2D MUSIC simulations of accreting stars \cite{Geroux2016}.

Finally, we note that \cite{Kunitomo2021} recently explored the effect of an increasing metallicity as the disc is being more structured and found that a more realistic chemical evolution of the accreted material could explain the structure of the modern Sun. However, this level of detail is beyond the scope of the present work and adds an extra parameter, therefore for now, we set the accreted material with the exact same composition as the central initial seed and do not change it as the disc evolves. 

\subsection{Angular momentum evolution}

The PMS evolution is highly dynamic and the angular momentum exchanges are mostly driven by the interactions with the disc. Thus, 1D evolutionary models cannot capture all the details of the accretion phase, only multidimensional MHD models may be able to estimate the dynamics and the overall physical trends which can then be input in 1D models. In the present case, we rely on the 2.5D PLUTO simulations by \cite{Pantolmos2020}, \cite{Ireland2021} and \cite{Ireland2022}. These simulations model the magnetised wind of a star accreting from a disc. They were able to extract some simplified expressions describing the fluxes of angular momentum leaving or coming onto the stellar surface as a function of physical parameters. 
During the star-disc interaction phase, the torque can be separated in three components. The first one is due to the accretion by the star of material incoming from the inner edge of the disc. A second one is due to a magnetised stellar wind, similar to that existing on the main sequence \citep{Matt2012,Reville2015a,Finley2018}, but with some modifications arising from the presence of the disc and the accretion of material. The last component of the torque is related to the large bubbles of plasma that leaves the star-disc system, due to a large scale magnetic reconnection close to the stellar surface. These so-called magnetospheric ejections (MEs) remain magnetically tied to the star and may either spin-up or spin-down the star depending on the accreting state. 

Indeed, the 3 components of the torque have conveniently been expressed so that only one parameter governs the change from one state to another : the ratio of the truncation radius $R_t$ of the disc to the co-rotation radius $R_{co}$ (i.e., where the Keplerian orbital rate is equal to the angular rotation rate of the star). In state 1, $R_t/R_{co}<0.433$ and the star is accreting from closer-in where the orbiting material has less angular momentum and thus weakening the spin-up torque. However, the MEs are also spinning up the star because the disc material brought by the magnetic field lines to the star comes from within the corotation radius where it spins faster than the surface.
State 2 can be seen as a transition stage, when $0.433<R_t/R_{co}<1$ the system becomes more complex. In the simulations, the stellar magnetic field can now connects to the disc beyond the co-rotation radius, and thus the MEs may spin up or down the star depending on the extent of the field line and the stellar rotation (with the transition from spin-up to spin-down occurring at intermediate values in the range of $R_t/R_{co}$).
Finally, state 3 is when $R_t/R_{co}>1$. At this point, the system enters the so-called propeller regime where the star alternates between being able to accrete or not on timescales of days to weeks. This is associated with fast rotation and/or strong magnetic field, leading to a very dynamical situation. In this state, the MEs are maximally efficient at extracting angular momentum while the accretion torque is largely decreased.

We rely on \cite{Ireland2022} to describe the loss and gain of angular momentum of the star with a surrounding disc.  We describe these torque formulations below, and we also adopt their fitted values for each of their formulation parameters (see table~\ref{tab:constant_torque}). Note however the change of sign compared to their formulation, in our case a positive torque add angular momentum to the star and therefore spins it up. In addition, the mass accretion rate $\dot{M}_{\rm acc}$ and mass-loss rate $\dot{M}_{\rm w}$ are given in terms of absolute values thus leaving aside the sign again.

\subsubsection*{Truncation radius}
        The torques depend on the inner radius of the disc, or in other words the truncation radius where the force exerted by the stellar magnetosphere disrupts the disc structure. Based on time-averaged values for all simulations, \cite{Ireland2022} gives an expression to compute the location of the truncation radius, 
        \begin{equation}
            \frac{R_t}{R_*} = {\rm min} \left[ K_{t,1} \Upsilon^{m_{t,1}}_{\rm acc}\; , \; K_{t,2} \Upsilon^{m_{t,2}}_{\rm acc}f^{m_{t,3}}\right]
        \end{equation}
        with $\Upsilon_{\rm acc} = \frac{B_*^2R_*^2}{4\pi |\dot{M}_{\rm acc}| v_{\rm esc}}$ as the disc magnetisation parameter \citep{MattPudritz2005,Bessolaz2008}. The constants $K_{t,1}, K_{t,2}, m_{t,1}, m_{t,2}$ and $m_{t,3}$ are set from \cite{Ireland2022}'s simulations. $f$ is the fraction of the equatorial surface velocity relative to the break-up speed. \LA{$v_{\rm esc}$ is the escape velocity and $B_*$ is the surface dipole strength.} The two cases allow for a transition between state 1 where the truncation radius varies as a simple power law in the magnetisation parameter, and state 2 and 3 where $R_t$ is much closer to the co-rotation radius and the centrifugal barrier starts to hinder the accretion. In case of high accretion rates (or more generally low disc magnetisation values), the truncation radius can approach the stellar surface. To be able to model the extreme case where the disc reaches the surface of the star we further impose the condition that $R_t \geq R_\star$.

\subsubsection*{SDI torque}
    The first torque associated to the SDI to consider is obviously from the angular momentum brought in by the material accreted from the truncation radius $R_t$, it follows 
    \begin{equation}
        \dot{J}_{\rm acc} \propto \dot{M}_{\rm acc}\left(GM_\star R_t\right)^{0.5}
        \label{Eq:torqueacc}
    \end{equation}
    with $\dot{M}_{\rm acc}$ the accretion rate, $G$ the gravitational constant, $R_\star$ and $M_\star$ the radius and mass of the star.
    
    The second SDI torque we will consider comes from the magnetospheric ejections (ME). They are an outflow coming from the reconnection and expansion of magnetic field lines coupling the star and the disc. Due to the interaction between the two systems, the mid-latitude regions of the dipole are highly dynamic and occasionally reconnect. This leads to some outflows that separate from the stellar magnetosphere and the system. Depending on the regime and the extent of the magnetic field with regard to the co-rotation radius, they are either spinning up or down the star \citep{ZanniFerreira2013,Gallet2019,Ireland2021,Ireland2022}.
    \cite{Ireland2022} found that the spin-up torque varies with the accretion torque such that 
    \begin{equation}
        \dot{J}_{\rm ME,up} \propto \dot{J}_{\rm acc} \propto \dot{M}_{\rm acc} (GM_*R_t)^{0.5}.
        \label{Eq:torqueMEacc}
    \end{equation}
    The spin-down torque associated with the MEs depends on the differential rotation between the stellar surface and the MEs connected to the disc. Assuming a dipolar magnetosphere, \cite{Ireland2022} found 
    \begin{equation}
        \dot{J}_{\rm ME,down} \propto -\frac{B_*^2R_*^6}{R_{\rm in}^3},
        \label{Eq:torqueME}
    \end{equation}
    with $R_{\rm in}=\textrm{max}(R_{\rm co},R_t)$.

    The torque associated directly with the SDI can be written as the sum of the accretion torque and the ME torque depending on the accretion state in which the system is at a given time. 
    According to \cite{Ireland2022}'s simulations, the net SDI torque can be expressed as
    \begin{equation}
    \dot{J}_{\rm SDI} = \left\{ 
    \begin{aligned}
        &K_{\rm SDI,1}\dot{M}_{\rm acc} (GM_*R_t)^{0.5} & \mbox{ if }& R_t < 0.433R_{\rm co} \\
        &K_{\rm SDI,1}\dot{M}_{\rm acc} (GM_*R_t)^{0.5} & \\
        &- K_{\rm SDI,2}B_*^2R_*^6/R_{\rm co}^3 &         \mbox{ if } &0.433R_{\rm co}<R_t < R_{\rm co} \\
        &-{K_{\rm SDI,2}B_*^2R_*^6/R_t^3}  &                 \mbox{ if }& R_{\rm co} < R_t.
    \end{aligned}
    \right.        
    \label{Eq:torqueSDI}
    \end{equation}

\subsubsection*{Spin-down torque : winds}
    At last, the magnetospheric accretion and the presence of the disc perturb the large scale magnetic field of the star and allows for more lines to be open, globally increasing the specific (per-mass-loss-rate) torque compared to a wind from a star without a disc. In addition, the hypothesis of accretion-powered stellar winds posits a wind mass-loss rate $\dot{M}_{\rm wind}$ that is a fraction of the accretion rate, generally inferior to $\sim$10\% \citep{MattPudritz2005}, but more likely with an upper limit around 1\% \citep{ZanniFerreira2011}.
    \begin{align}
    \begin{split}
        \dot{J}_{\rm wind} = - K_{A,1}^2(\alpha\pi  K_{\Phi})^{4 m_A} |\dot{M}_{\rm acc}|\sqrt{GM_\star R_\star} \left(\frac{\dot{M}_{\rm wind}}{\dot{M}_{\rm acc}}\right)^{1-2 m_A}\\ 
        f^{1+4m_Am_{\Phi,2}}\left[1+\left(\frac{f}{ K_{A,2}}\right)^2\right]^{- m_A}\left(\frac{R_t}{R_\star}\right)^{4 m_Am_{\Phi,1}}
    \end{split}
    \label{Eq:torquewinds}
    \end{align}
    with $f$ the ratio of the \LA{stellar} surface \LA{rotational} velocity to the \LA{break-up speed} and $K_\textrm{A,1}$, $K_\textrm{A,2}$, $K_{\Phi}$, $m_A$, $m_{\Phi,1}$, and $m_{\Phi,2}$ are constant again calibrated on MHD simulations and presented in table~\ref{tab:constant_torque}. In this work we will only consider a surface dipole as it is the geometry that has been chosen for \cite{Ireland2022}'s simulations, thus $\alpha = 2$.
            
\begin{table*}[]
    \centering
    \begin{tabular}{c c c c c c c c c c c c c c}
        $K_{\rm SDI,1}$ & $K_{\rm SDI,2}$ & $K_{A,1}$ & $K_{A,2}$ & $K_{t,1}$ & $K_{t,2}$ & $K_{\phi,1}$ & $m_A$ & $m_{\phi,1}$ & $m_{\phi,2}$ & $m_{t,1}$& $m_{t,2}$& $m_{t,3}$ & $\alpha$\\
        &&&&&&&&&&&&& \\
        \hline
        &&&&&&&&&&&&& \\
        0.909&0.0171&0.954&0.0284&0.772&1.36&1.62&0.394&-1.25&0.184&0.311&0.0597&-0.261&2.0 \\
        &&&&&&&&&&&&& 
    \end{tabular}
    \caption{Parameters used for the SDI torque}
    \label{tab:constant_torque}
\end{table*}  

\begin{table}[]
    \centering
    \begin{tabular}{c|c c}
         & Reference & Explored range \\
         \hline
         $\xi$ & 0.10 & $0.02 - 0.50$ \\
         $m_{ke}$ & 0.02 & - \\
         $\dot{M}_{\rm acc, ini} (10^{-5}M_\odot.yr^{-1})$ & $1.15$ & $0.75-1.66$ \\
         $n$ & 1.5 & $1.2-2.0$ \\
         $M_{\rm ini} (M_\odot)$ & 0.01 & - \\ 
         $\alpha_{MLT}$ & 2.173 & - \\
         $X_D [ppm]$ & 28 & 10 - 50\\
         $\dot{M}_{\rm w}/\dot{M}_{\rm acc}$ & 0.05 & $0.01-0.05$\\
         $B_*$ (G) & 1000 & 750 - 2000 \\
         $\tau_D$ (Myr)& 10 & -
    \end{tabular}
    \caption{Reference model and explored parameter range}
    \label{tab:Reference}
\end{table}

\subsection{Initial conditions and reference model}
    \subsubsection{Initial model}
        We obtain the initial model by removing the external layers of a 0.1M$_\odot$ polytrope, and then relax the remaining seed for about a thousand years with STAREVOL. We verify that the resulting seed was not dependent on these assumptions and end up with an initial seed of 0.01M$_\odot$ for a radius of about 1.5 R$_\odot$ similarly to \cite{Stahler1980}, \cite{Hosokawa2011} and \cite{Kunitomo2017}. Our initial seed is a little larger than the second Larson core \citep[\textit{e.g}][]{Larson1969,Vaytet2013} and correspond to a relatively high initial entropy case \citep{Baraffe2009,Kunitomo2017}.
        Note that 3D MHD numerical simulations of star forming cores seem to indicate a much larger radius for this mass and at this stage \citep{Bandhare2020}, indicative of an extremely high accretion rate with very little radiative losses. The accretion history as well as the physics in these models are quite different from what we are using in this study, in addition it has not yet been confirmed by observations and we thus kept a smaller initial radius.
        
    \subsubsection{Accretion characteristics and assumptions for the reference case}
    \label{sec:accretion}
\begin{figure}
            \centering
            \includegraphics[width=0.4\textwidth]{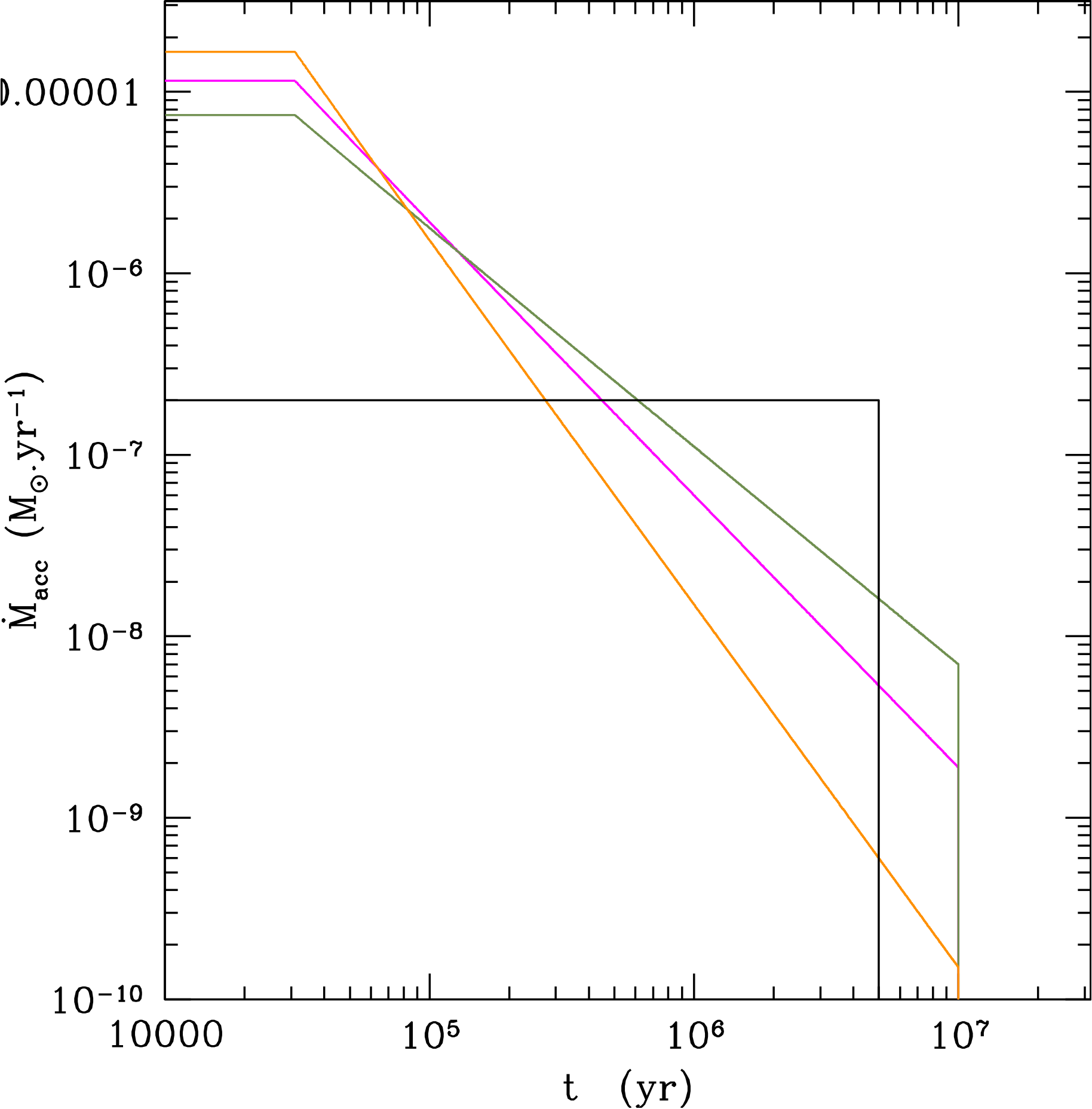}
            \caption{Evolution of the accretion rates as a function of time for the four cases studied in sect.~\ref{sect:stdacc}. The orange, magenta (reference), and grey tracks show the power-law decays with indexes 2.0, 1.5, and 1.2, respectively. The Black line shows the constant accretion rate at $\dot{M}=2.10^{-7}\dot{M}_\odot$}
            \label{fig:Acc_hist}
        \end{figure}

        We set a reference case in order to test the different parameters of the model. All the inputs are summarised in table~\ref{tab:Reference}.  
        In this work, for simplicity and to focus on the effect of the torque on the rotational evolution, we chose to use an analytical form for the accretion rate. We acknowledge more realistic modelling where the accretion rates come from accretion disc simulations by \textit{e.g.} \citet{Gehrig2022}  \citep[See also][]{Baraffe2017,JensenHaugbolle2018,Miley2021,Kunitomo2021}. In these works the accretion histories are highly variable (episodic) and create very bumpy evolutionary paths in the HR diagram. In the current case, the evolution of the accretion rate is smooth and defined by a simple function allowing for well defined path. 
        We start the evolution with a strong constant accretion rate $\dot{M}_{\rm acc,ini}$ for $3.10^4$yr. After that time, the accretion rate decays as a power law with index $n$ based on the observation in young accreting stars. We thus have 
        \begin{align}
            \dot{M}_{\rm acc}(t) = \dot{M}_{\rm acc,ini} \left(\frac{{\rm max}(3.10^4,t)}{3.10^4}\right)^{-n},
        \end{align}
        where $n$ has typical values between 1.2 and 2. We set a disc duration time of 10 Myr for all models, after which time the accretion rate drops to zero.
        For each value of $n$ we explored, we set the value of $\dot{M}_{\rm acc,ini}$ such that a total of 99\%$M_\odot$ is accreted by the end of the accretion phase, bringing the final stellar mass to $1 M_\odot$. 
        In each case, $\dot{M}_{\rm acc,ini}$ has a value around $\sim$10$^{-5} M_\odot$yr$^{-1}$ (see Table \ref{tab:Reference}).
        The three decaying accretion rate histories are displayed on figure~\ref{fig:Acc_hist}. In addition we also present a case of constant accretion at $\dot{M}=2.10^{-7}M_\odot$yr$^{-1}$ for 5 Myr.
        For accounting mass added to the star, we neglect the mass loss from stellar winds.
        
        The composition of the disc may evolve with time and become quite different as the gas-grain ratio changes with the disc structure. While the gas is first being accreted or blown away by the winds, the heavier elements have condensed into grains and start to form small pebbles, they can then chemically enrich the accreted material and creates long-time effects on the stellar structure \citep{Kunitomo2018}. In the current case, we keep a constant chemical composition identical to the initial proto-stellar seed. The initial amount of deuterium is set at 28 ppm which is close to the value obtained from meteorites \citep{AsplundGrevesse2009}. 
        
        Following \cite{Hartmann1997}, we assume that most of the entropy is radiated away as the material reaches the surface, thus the specific entropy of the accreted material is \LA{about} the same as the stellar surface.
        The accretion efficiency parameter $\xi$ is fixed at 0.1 for the reference model which means that only 10$\%$ of the heat of the infalling material is absorbed by the star. This corresponds to a relatively cool accretion. \cite{Hartmann1998} and \cite{Kunitomo2017} showed that this value was able to explain the broadening of the sequence observed in \LA{the HR diagram of } very young open clusters.
        Finally, the energy and accreted mass are deposited in the upper 2$\%$ of the star (See Sect.~\ref{Sect:FormAcc} before being transported by the convective motions.

\section{Stellar evolution with accretion and no rotation}
\label{Sect:nonrot}
\subsection{Reference case}
\label{Sect:nonrot_ref}
        \begin{figure}
            \centering
            \includegraphics[width=0.45\textwidth]{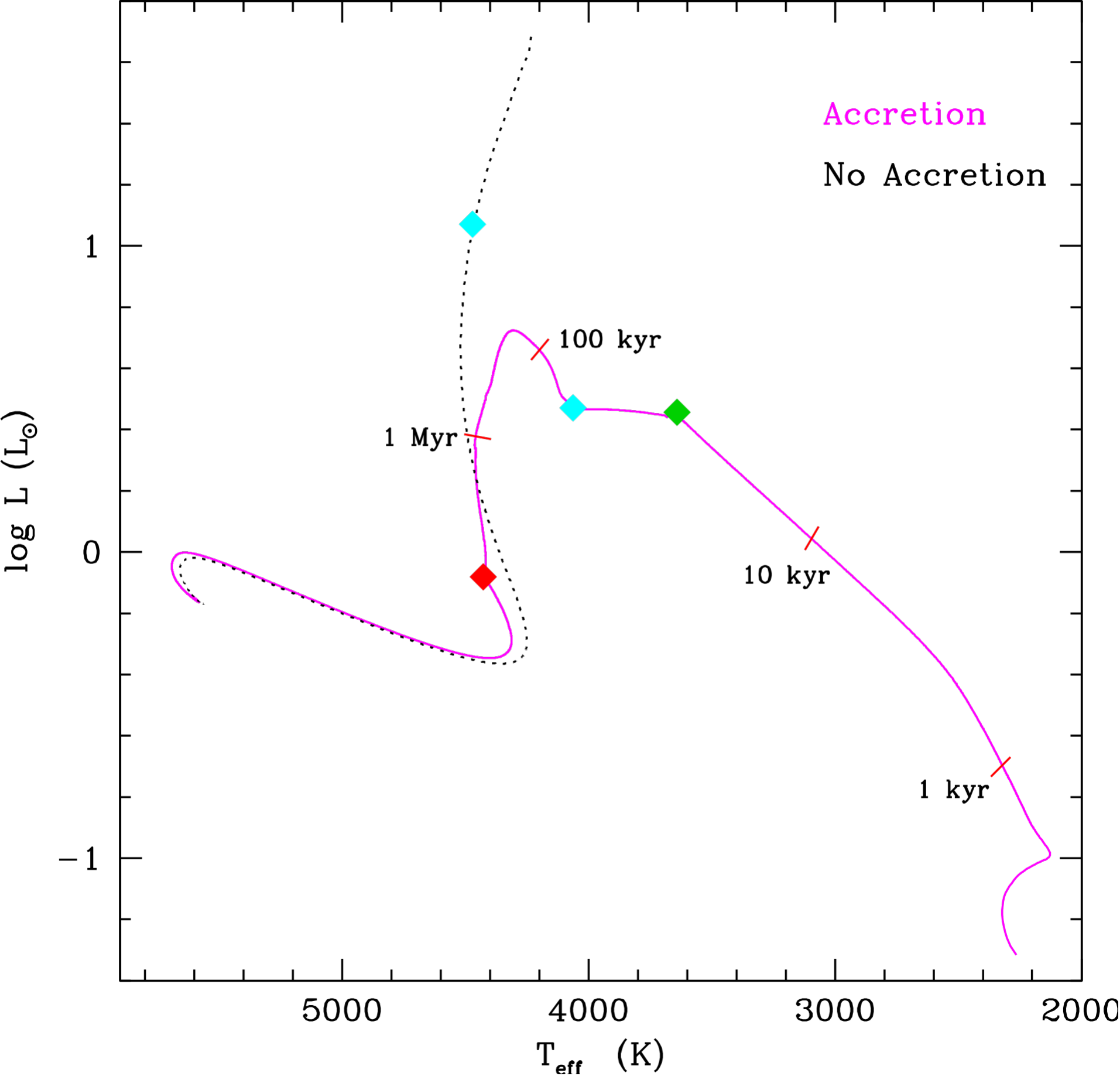}
            \caption{Hertz\LA{s}prung-Russel diagram of a 1M$_\odot$ pre-main sequence star starting from 1M$_\odot$ (without accretion, black) and from 0.01M$_\odot$ with accretion, our reference case (magenta). The green, red, and cyan diamonds indicate the end of the constant high accretion rate, the end of the accretion phase, and the onset of deuterium burning, respectively.}
            \label{fig:HR_ref}
        \end{figure}
        
        Figure~\ref{fig:HR_ref} shows the evolution of our reference case compared to a classical model for solar-mass PMS evolution. The starting point is obviously very cold and despite its relatively large radius it remains faint compared to the solar mass evolution.
        During the early evolution, the accretion rate is very strong and the stellar structure quickly reacts to the large amount of accreted energy by strongly increasing the model's radius from 1.5 to more than 2.5 R$_\odot$ in a few iterations. 
        Figure~\ref{fig:Non_rot_Deut} shows the evolution of stellar radius for the reference model (magenta line) compared to the classical (non accreting) model (red dashed line), starting from an age of 10~kyr.
        The radius (and the temperature), and thus the luminosity keep increasing as the mass increases to reach about 4.0 R$_\odot$ (3600K) before the accretion rate starts to decrease at thirty thousand years. 

        Around $7.10^4$yr, the temperature has increased enough in the inner region (around 10\% in mass) to ignite the burning of deuterium. As the reactions onset, a convective layer develops deep in the star and reconnects with the upper convective envelope, suddenly increasing the radius and the luminosity of the star at $\log(T_{\rm eff})\sim3.6$. At this point, the star is about half a solar mass and is fuelled by the accreted deuterium. It continues its evolution in the HR diagram (Fig.~\ref{fig:HR_ref}) close to the path of non-accreting PMS model of the same mass.
        Note however that in our case, the deuterium burning still contributes to the stellar luminosity all the way to the end of the SDI phase, thereby increasing the radius of the star to the point of being slightly larger than the non-accreting case beyond about 1 Myr.

\subsection{Deuterium abundance}
\begin{figure}
        \centering
        \includegraphics[width=0.45\textwidth]{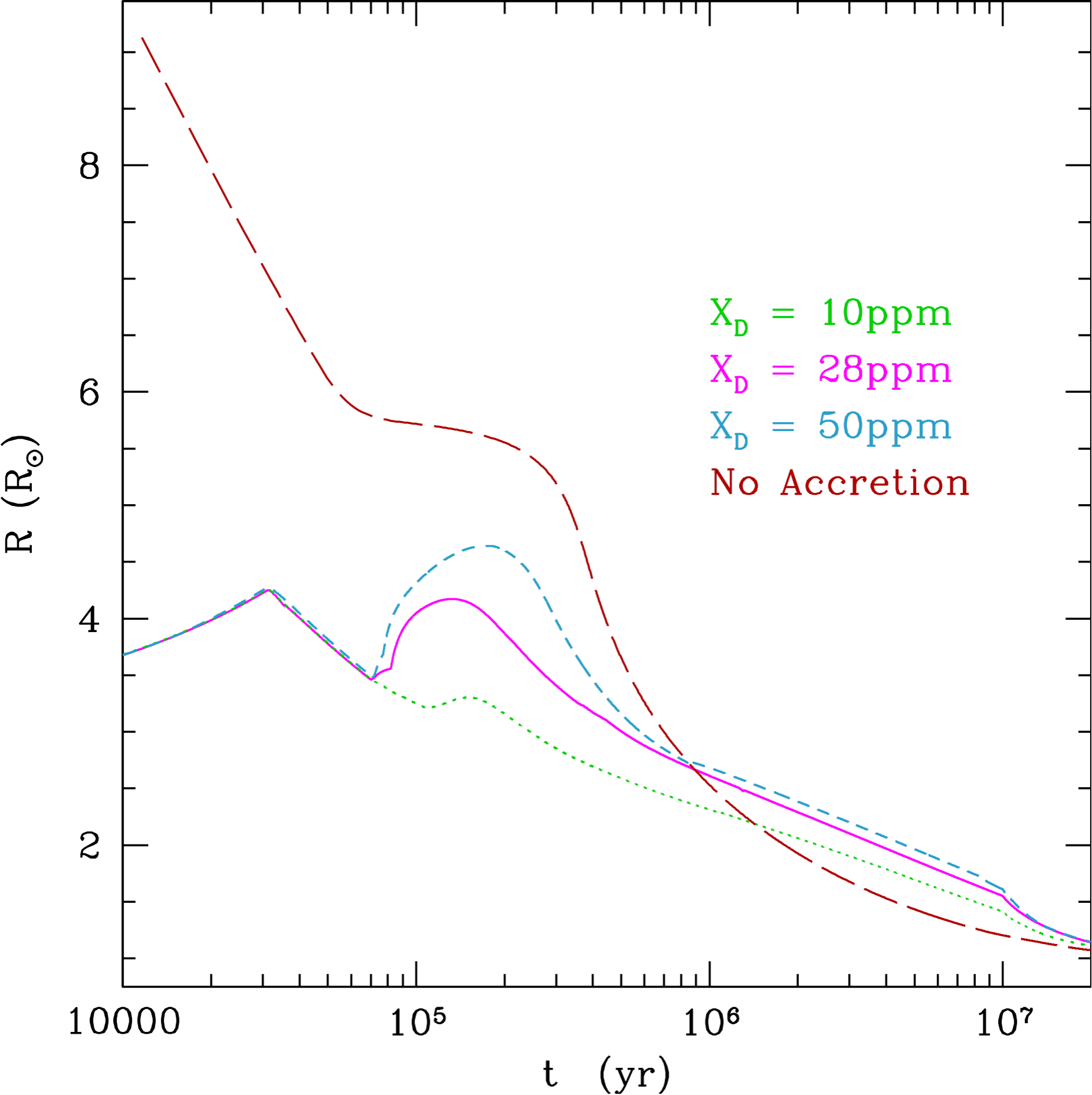}
        \caption{Radius as a function of time for models accreting material with $X_D=10$ppm (green), $X_D=28$ppm (magenta), and $X_D=10$ppm (blue), and a non-accreting models with an initial abundance of $X_D=28$ppm (dark red).}
        \label{fig:Non_rot_Deut}
    \end{figure}

    As of today, the initial amount of Deuterium present in the Sun is still poorly constrained. It cannot be tracked from today's abundance and the spectroscopic analysis of meteoritic rocks give a value that only constrains the region of the disc in which the meteorite material was formed.
    Since during the PMS, deuterium is the only element that can be burnt to counteract the gravitational contraction, it is crucial to test its influence on the evolution. If the amount of deuterium is fixed from the starting point its combustion will last a few hundred thousand years. However, if it is being constantly refuelled by newly accreted material, it has a long lasting effect on the structure.
    On figure~\ref{fig:Non_rot_Deut}, we show the radius evolution for different deuterium mass fractions from 10ppm to 50ppm.   
    We remind the reader that for this work we assumed the same composition for the initial seed as for the accreted material.
    As expected, the evolution is the same in all cases before the onset of Deuterium burning at about sixty thousands years.
    Then, most of the luminosity is coming from the nuclear reactions and since most of the deuterium is consumed, the larger the mass fraction of deuterium, the larger the luminosity and radius. For example, at 2.10$^5$ yr, the high deuterium model ($X_D=50$ppm) radius is larger by a factor 1.4 than the low-deuterium case ($X_D=10$ppm). During the later stages (1-10 Myr), the difference is much lower but remains qualitatively the same---the larger the fraction of deuterium fuelling the star, the more the contraction is limited.  Note that this radius evolution differs from figure 2 of \cite{Kunitomo2017}, essentially because they considered a cold case ($\xi=0$) while we have a warm accretion ($\xi=0.1$).

\subsection{Accretion history}
    \label{sect:stdacc}
    We mentioned in the introduction that the accretion rate is quite variable along the evolution \citep{BC10} and here we only reproduce some global trends. As the disc becomes thinner and more structured, the accretion rate also becomes generally lower. The slope with which the decays happen is not strongly constrained by observations. Some studies use isochronal ages to estimate the evolution of accretion rates with time \citep[e.g.][]{CoG2012}, but at this stage of the evolution the uncertainties in age are very large, up to a factor more than 2 per open clusters \citep[e.g.][]{Belletal2013, Cao2022}, and even more if we consider an internal age spread \citep{Getman2018}.
    We consider four different simple accretion histories : one constant accretion rate at $\dot{M}_{\textrm{acc}} = 2.10^{-7}M_\odot.yr^{-1}$ and three power-law decaying accretion rates of the form $\dot{M}_\textrm{acc} = \dot{M}_{\textrm{ini}}.10^{-n}t$ with $n=[1.2, 1.5, 2.0]$. In order for all models to reach a mass of one solar mass by the end of the accretion phase, we shift the initial accretion rate, as described in \S\ref{sec:accretion} (see Tab.~\ref{tab:Reference} and Fig.~{\ref{fig:Acc_hist}).
    Finally we use a non-accreting model of 1$M_\odot$ for comparison purpose.         \LA{In that case the evolution starts with a one solar mass polytrope similar to what is typically used in rotation evolution models. Note that the initial radius of ($\sim10R_\odot$) is larger than what is expected on the birthline \citep{StahlerPalla2005}, although it makes the corresponding initial rotation rate slightly higher, it remains one order of magnitude lower than other models.}
    
    On figure~\ref{fig:Non_rot_Macc}, we show the radial evolution of models with each of the varied accretion rates, compared to a 1M$\odot$ non-accreting case.
    The first thing we notice is that the higher the initial accretion rate, the larger the stellar radius, likely because of the larger amount of energy deposited by the accretion in the upper layer of the star. Even though the $n=1.2$ case has the higher accretion rate already at 0.15 Myr, the total mass remains much smaller and at this point of the evolution the Deuterium burning has taken over the accretion in terms of input of energy. Thus, for the trend to reverse, we need to wait for another $\sim$1 Myr when the radius of the higher accretor is back to be the largest ($n=1.2$ case).
    The constant accretion rate case is a little special. It is comparatively very low at first with a mass accretion rate almost 2 orders of magnitude lower than the other models, and becomes between 1 to 3 orders of magnitude larger during the later evolution. This leads to a very cold and compact evolution early on because of the low amount of mass accumulated. The star then accumulates 80$\%$ of its mass from 1 Myr to the end of the disc phase, and thus keeps gaining energy from accretion and fresh deuterium to burn, and the star puffs up accordingly.

\begin{figure}
    \includegraphics[width=0.48\textwidth,right]{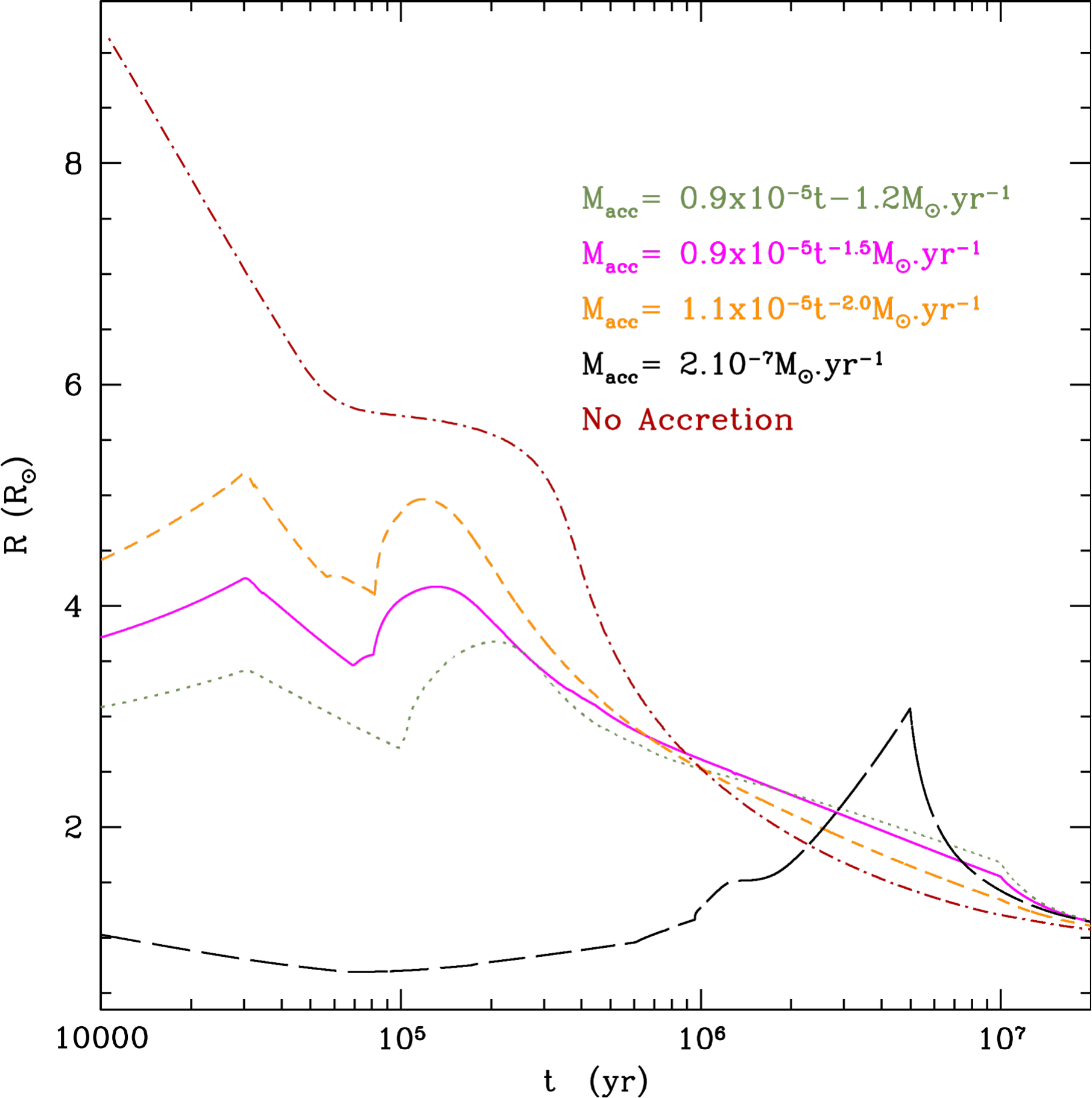}
    \caption{Evolution of the radius for the four different accretion rates histories as well as the non-accreting case.}
    \label{fig:Non_rot_Macc}
\end{figure}

\subsection{Accreted energy}
The entropy of the accreted material was shown to drastically affect the evolution \citep[\textit{e.g.}][]{Hartmann1998}. A few elements have to be considered here. First, the material comes in with its own entropy which is generally \LA{similar to the stellar surface value.}
This is what we will call the warm accretion (cold if entropy is null). Thus, we consider that most (but not all) of the kinetic energy of the accreted material has been radiated away before being deposited in the star. Furthermore, in this work we consider that the accreted heat is only injected in the upper layers $m_{ke}$ of the star (in fraction of mass) before being transported by the convective motions to the deeper layers.
On figure~\ref{fig:Non_rot_alph} we present the radius evolution for three cases from a very cool accretion ($\xi=0.02$, blue), to a hot accretion ($\xi=0.50$, cyan) with an intermediate warm accretion (reference case; $\xi=0.10$, magenta).
The lower heat content reduces the inflation during the early evolution, while the hot accretion model expands to very large radius, even approaching the non accreting case.
\cite{Kunitomo2017} pointed out that the heat injection is likely the most important parameter during the first growing phase. Here, we confirm that the differences are quite striking, even more than in the cases of different accretion histories or deuterium contents (for the particular ranges of parameters explored here). 
However, after about 1 Myr, the structures have converged despite a very different initial evolution.

\begin{figure}
    \centering
    \includegraphics[width=0.45\textwidth]{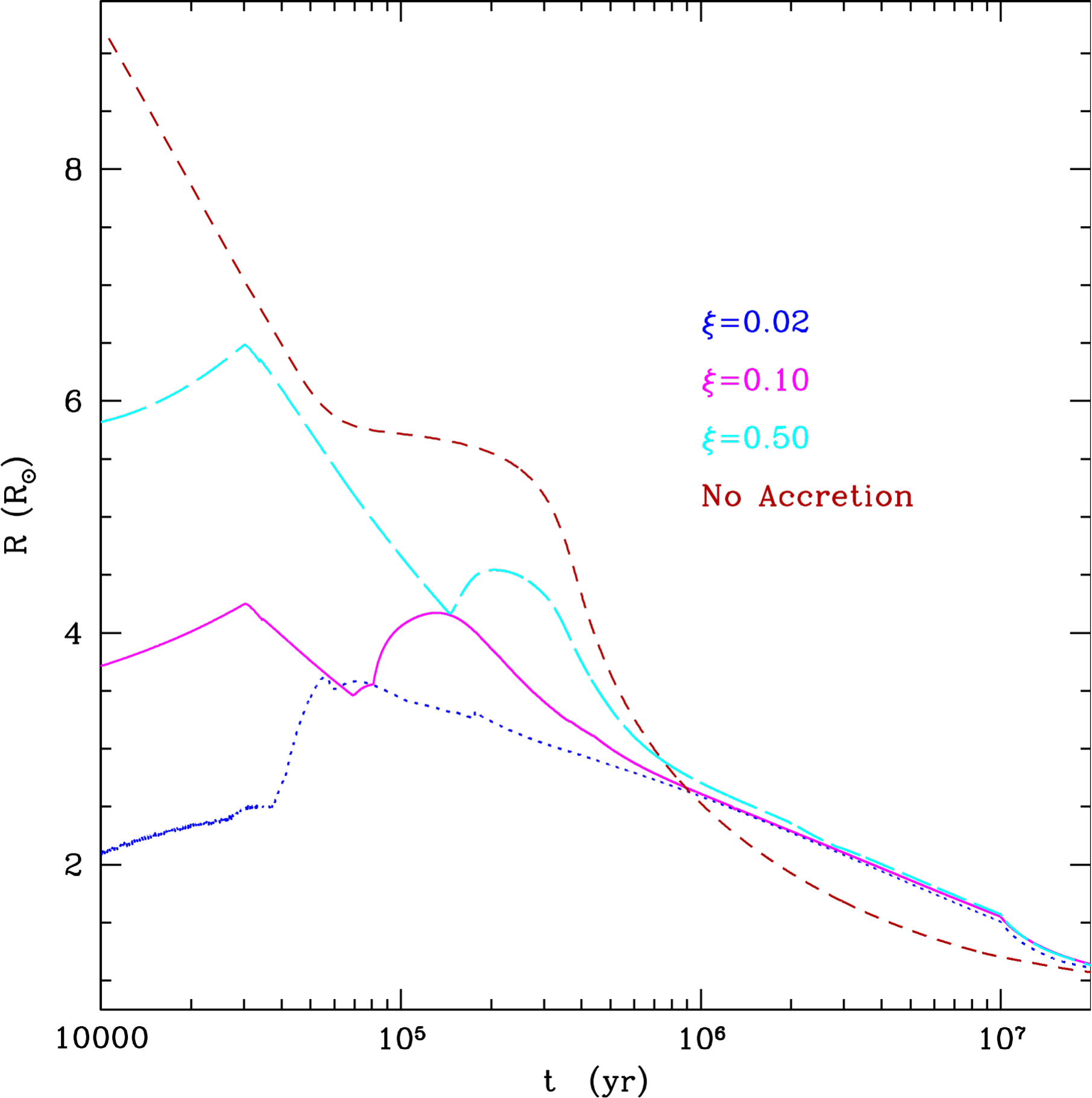}
    \caption{Evolution of the radius for three models with different entropy content in the accreted material}
    \label{fig:Non_rot_alph}
\end{figure}

\section{Evolution with accretion and rotation}

The novelty of this work is to consider both at the same time and in a semi-consistent way the rotation and the stellar structure evolutions during the accretion phase. We showed in the previous section that the stellar structure is strongly affected by the accretion of material and that different assumptions on the accretion properties lead to widely different evolution of the structure and thus of the moment of inertia.
In this section we look at the evolution of the surface rotation of such models when their angular momentum content is driven by the exchange with the surrounding proto-planetary disc.
To compute the SDI and wind torques we need to set the properties of the magnetic field as well as the mass-loss rate. For the reference model, we use a 1kG dipole and a mass-loss rate equal to 5$\%$ of the mass-accretion rate. We then proceed to broaden this range of magnetic field strength and explore lower mass-loss rate ratio which may be closer to what we can expect.

Note that the initial rotation period is often a parameter with a very high uncertainty. However, in the case of models accreting mass and angular momentum, varying the initial velocity does not make a difference because the moment of inertia of the initial seed is so small, compared to that of the final object. Indeed, the amount of angular momentum of the initial seed is negligible compared to the angular momentum brought in by the SDI torque (see fig~\ref{fig:Alltorques}). We thus arbitrarily chose an initial surface velocity of 4km.s$^-1$ for the initial seed which corresponds to a period of about 20 days. 

\subsection{Reference model}
\subsubsection{Evolution of the different torques}

\begin{figure*}
    \centering
    \includegraphics[width=0.7\textwidth]{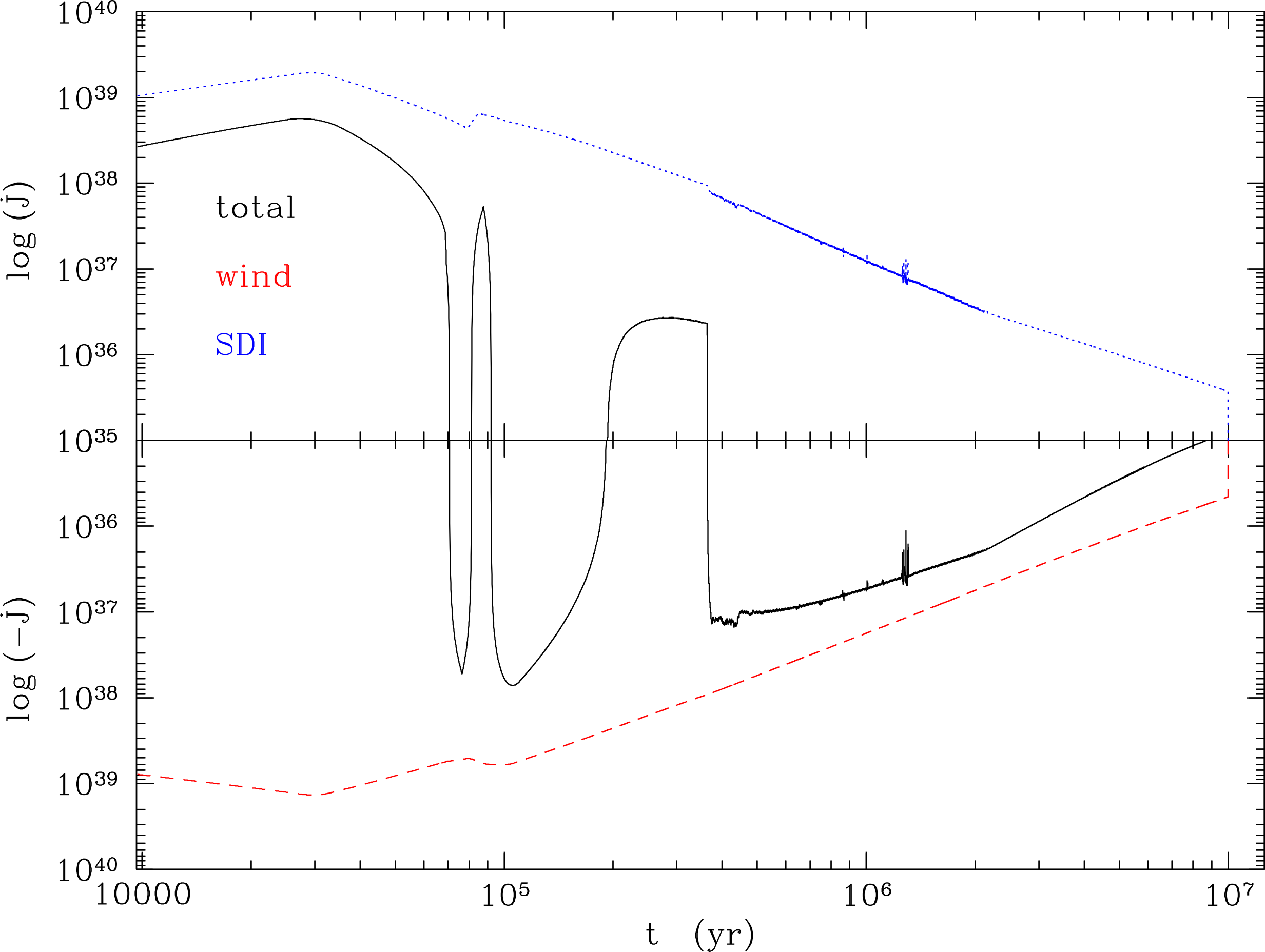}
    \caption{Torques guiding the evolution of the stellar angular momentum as a function of time for the reference model in g.cm$^2$.s$^{-2}$. The upper (lower) part of the diagram shows a spin-up(-down) torque applied on the star. The total torque is the solid black line, the stellar wind torque is plotted in dashed red line, and the torque due to the SDI interaction (sum of the accretion and ME torques) is in dotted blue.}
    \label{fig:Alltorques}
\end{figure*}

\begin{figure*}
    \centering
    \includegraphics[width=0.7\textwidth]{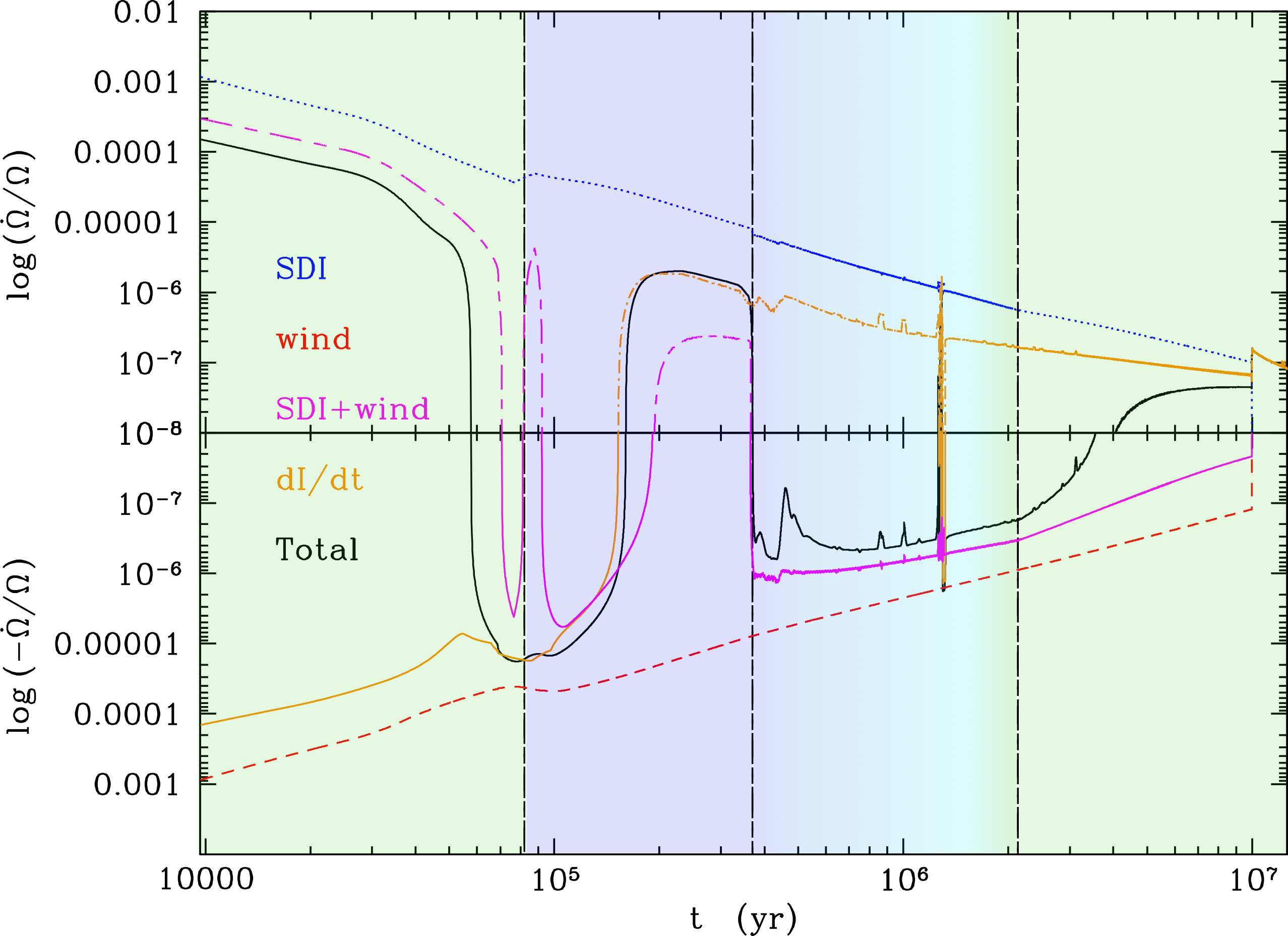}
    \caption{Variation of the rotation rate associated with each torque in fraction of the current rotation rate per year. The upper (lower) part of the diagram shows a spin-up(-down) of the star. The actual change of the rotation rate is shown in solid black. The one associated with the stellar wind and the SDI torque are plotted in dashed red and dotted blue lines respectively. The variation due to a change of the moment of inertia is shown in orange. The background colours indicate the state of the star-disc interaction with state 1 and 2 shown in purple and green, respectively. The light blue region with a gradient indicates a long-lived phase when the system remains close to the state-1-2 transition and switches frequently between the two states. This model does not reach state 3.}
    \label{fig:AllDOmega}
\end{figure*}

The evolution of the total torque is rather complex, with intermittent phases of net positive or net negative torque, as the system evolves.
Figure~\ref{fig:Alltorques} shows the torques exerted on the star by the SDI (\textit{i.e.} the sum of the accretion and the ME's torque) and the magnetised stellar winds in the reference case, as a function of time. 
We see that torques change by more than four orders of magnitude between the beginning and the end of the disc phase. Nevertheless, the SDI torque approximately mirrors the enhanced stellar wind torque, and therefore the total change in angular momentum is much smaller than either of the two torques and is alternatively led by one of the two components.
Figure~\ref{fig:AllDOmega} helps to visualise the effect of each torque on the stellar spin evolution. Since the spin rate is also affected by the change of moment of inertia, we added the relative variation of $\Omega$ due to the change of inertia over the course of the disc-coupling phase. The latter can be compared to the rotational change due to the sum of the two magnetic torques (SDI+winds, magenta line on figure~\ref{fig:AllDOmega}).
It is clear that the rotational evolution of the star is quite complex, undergoing intermittent phases where the spin evolution is dominated by the net torque, dominated by the changing moment of inertia, or set by a combination of these factors.

At first, when the accretion rate is very strong $(\sim 10^{-5} M_\odot .yr^{-1})$, the star is gaining angular momentum very rapidly. At this stage, the truncation radius of the disc is limited by the surface of the star, and the star is thus accreting through boundary layer accretion. 
As the star rotates more rapidly and the accretion rate decreases, both the wind and the MEs reduce their action. 
At the same time, the radius of the star expands very rapidly because of the energy brought in by the accreted material.
In this situation, the mass as well as well as the radius of gyration \citep{Rucinski1988} strongly increases. 
Note for example that despite the radius shrinking from 30 to 60kyr as the accretion rate start decreasing, the moment of inertia still grows due to the expansion of the convective region, thus overall, the stellar moment of inertia still grows and acts to spin down the star as visible on figure~\ref{fig:AllDOmega}.
Around 7.10$^4$ yr the deuterium burning starts in the deeper layer of the star (see section~\ref{Sect:nonrot_ref}).

Soon after, around 8.10$^4$ yr, the deep convective layer connects to the base of the convective zone, bringing the freshly accreted deuterium down to the regions where it can be burnt. 
The release of energy hence puffs up the star. As a consequence, the star goes to state 1 and the SDI torque (blue on figures~\ref{fig:Alltorques} and~\ref{fig:AllDOmega}) becomes a pure spin-up, thus losing the spin-down torque associated to the MEs. 
Note that during the whole time the star spent in state 1, the spin-evolution is dominated by the change of moment of inertia (dI/dt is higher than the SDI+wind torque). Indeed, despite the stronger SDI torque (blue line on fig~\ref{fig:AllDOmega} and~\ref{fig:Alltorques}), the sudden increase of radius keeps the star in a spin-down state for about a hundred thousand years.
As the radius inflation stops, the moment of inertia starts to decrease and spins up the star with a little help from the accretion torque.

Finally, between about $4.10^5$ yr and $2.10^6$ yr, the truncation radius stays very close to 0.433 times the co-rotation radius, which demarks the boundary between states 1 and 2. The star oscillates between state 1 and 2 on a timescale of the order of a hundred years at first up to a thousand years as the accretion rate decreases, and the evolutionary time-step then becomes very small to follow this phase. It results an overall loss of angular momentum from the star. 
After 2 Myr, the star is clearly in state 2 as the mass accretion rate becomes smaller and smaller and the MEs start to efficiently remove angular momentum from the star. As seen on figures~\ref{fig:Alltorques} and~\ref{fig:AllDOmega}, the combined SDI and wind torques then spin down the star, but the mild contraction of the star has the star slowly spinning up from about 4 Myr to the end of the disc-coupling phase at 10 Myr.

Figures~\ref{fig:Alltorques} and~\ref{fig:AllDOmega} only display the evolution of the torque and spin evolution of the reference case. The other cases with different physical assumptions discussed in Sect.~\ref{Sect:nonrot} all show qualitative similarities. Each shows varying epochs of spin up and down due to varying balances between the torques and the structure evolution. For the sake of brevity, we do not discuss them here.

\subsubsection{Comparisons to observations}
\cite{Covey2005} published the projected surface velocity ($v\sin i$) for 38 Class I objects.
However, the exact evolutionary stage is not well constrained. Their $v\sin i$ range from 10 to 60 km.s$^{-1}$ with a median projected velocity of 38 km.s$^{-1}$.

The lack of information on the rotation of Class I young stellar object limits the possibility to constrain our model on the earliest phases of the evolution.
Thus, to test our models against observations, we select the younger clusters of \citet[][also present in \citealt{Gallet2019}]{GB15}  and NGC2264 from \cite{Venuti17} as a comparison point (see table~\ref{tab:obs}). Their rotation period distributions are similar, each exhibits a broad range of rotation rates within a similar range, between $\sim$1 and $\sim$10 days, therefore the large uncertainty on their age does not strongly affect their quality as indicators.

On figure~\ref{fig:obsrot_D_E_Macc}, we show the rotation period evolution for all models with different input physics. In the background, we display the rotation period distribution of 1$\pm$0.1M$_\odot$ stars in six young clusters from the ONC (1.5Myr) to hPer (13Myr), as listed on table~\ref{tab:obs}. In addition we show the median, 25th, and 90th percentile for the clusters with a sufficient number of datapoint \citep{GB13,GB15,Gallet2019}.

\begin{table}[]
    \centering
    \begin{tabular}{c|c c}
       Name & Age (Myr) & Reference \\
         \hline
         ONC & 1.5 & \cite{RodriguezLedesma09} \\
         NGC6530 & 2 & \cite{HendersonStassun12} \\
         NGC2264 & 3 & \cite{Venuti17} \\
         CepOB3b & 4 & \cite{Littlefair10} \\ 
         NGC2362 & 5 & \cite{IrwinBouvier2009} \\
         hPer & 13 & \cite{Moraux13}
    \end{tabular}
    \caption{Ages and reference for the relevant open clusters and associations}
    \label{tab:obs}
\end{table}

First, the one solar mass non-accreting model undergoing only a pure contraction predicts a very steep spin-up during the entire PMS evolution. We start this model with an extremely slow spin rate, just so that the rotation period at the ZAMS would be reasonable, for illustration purpose.
On the contrary, none of the accreting model undergo such a strong, systematic spin-up between 0.1Myr and the end of the disc phase. 

We saw in section~\ref{Sect:nonrot} that accreting models start with a much lower radius and thus contract much less, also because of the continuous refuelling of deuterium by the accreted material. The slower contraction prevents a strong spin-up that would naturally come from the global conservation of angular momentum.

The reference model (in magenta on fig~\ref{fig:obsrot_D_E_Macc}) falls close to the median of the data. There is a little offset ($\sim$0.2 dex) towards a faster rotation depending on the considered cluster.
We then vary the same physical ingredients as in section~\ref{Sect:nonrot} as well as the dipole magnetic strength and the efficiency of the accretion powered stellar winds. 

\subsection{Exploring varying accretion assumptions}
Figure~\ref{fig:obsrot_D_E_Macc} present the rotation period evolution during the SDI phase for models with different accretion physics.

The top panel shows the evolution for 3 different amount of deuterium in the accreted material. Deuterium does not directly affect the SDI torque nor the wind torque. However, as we saw in the previous section, it inflates the star while burning, and thus increases the radius and the moment of inertia. As a consequence, as soon as the deuterium burning onsets (in these models around seventy thousands years), the rotation rate rapidly decreases at a rate dependent on the amount of deuterium accreted by the star.
This establishes a difference in rotation rate between the models that lasts as long as the star is accreting, although its effects lessen as the accretion rate decreases over time.

In the same fashion, the middle panel compares rotation period evolution with different heat injection by the accreted material. This parameter does not affect directly the torque but changes again the radius of the star. However this time the change in radius happens from the start of the evolution. A hot accretion leads to a large radius very early in the evolution, which limits the spin up during this phase because 1) the moment of inertia increases and 2) the larger radius and rotation period leads to a large ratio $R_*/R_{co}$ during the early evolution.
However, between 1 and 2 Myr, the tracks converge toward the same rotational evolution. At this point, the accretion energy becomes minimal compared to energy brought by the deuterium burning and the gravitational contraction, thus their structures become very similar (see figure~\ref{fig:Non_rot_alph}).
The injection of non-thermalised material therefore only changes the evolution during the first Myr of evolution and does not seem to play at role at later time, at least in our reference setup.
This may suggest that the stellar rotation is driven to a quasi-equilibrium spin rate by the various torques and structural evolution on a short timescale, and that this equilibrium is independent of their past evolution. 

The last panel shows the rotational evolution for different accretion histories. All models accrete the same amount of mass but not at the same time. In this case, the change in accretion rate modifies both the moment of inertia and the torques themselves. Thus the trends are not monotonous anymore. For example, the grey, magenta and orange tracks represent models with an accretion rate decaying more (orange) or less (grey) quickly from the end of the early strong accretion phase to the end of the disc-coupling phase at 10 Myr. We saw in the previous section the radius of the orange (grey) model with $\dot{M}\propto t^{-2}$ ($\dot{M}\propto t^{-1.2}$) is larger (smaller) at the beginning of the evolution. The larger expansion leads to a larger rotation period compared to the intermediate reference case. However, the lower accretion rate at later ages leads to a lower torque which allows the contraction to more strongly spin up the star as the model reaches the end of the disc-coupling phase.
The unrealistic constant accretion model ($\dot{M} = 2.10^{-7}M_\odot.$yr$^{-1}$) has a wild rotational evolution which reflects its structural evolution. The total mass of this model remains much lower than the reference case until the end of the disc phase (at 5Myr for this specific model).  As a consequence, the radius remains much smaller even though the star gains angular momentum, and thus spins up to very fast rotation before 1Myr. At this time, the model slowly starts to expand as the mass increases, and the accreted energy and deuterium are injected at a medium but steady rate. The model ends up being the largest at the end of the disc phase (at 5~Myr for this model) but with a similar rotation period as the others. The ensuing contraction towards the zero-age-main sequence makes it the fastest  rotator at the end of the PMS by a factor of five.

\begin{figure}[h]
    \centering
    \includegraphics[width=0.42\textwidth]{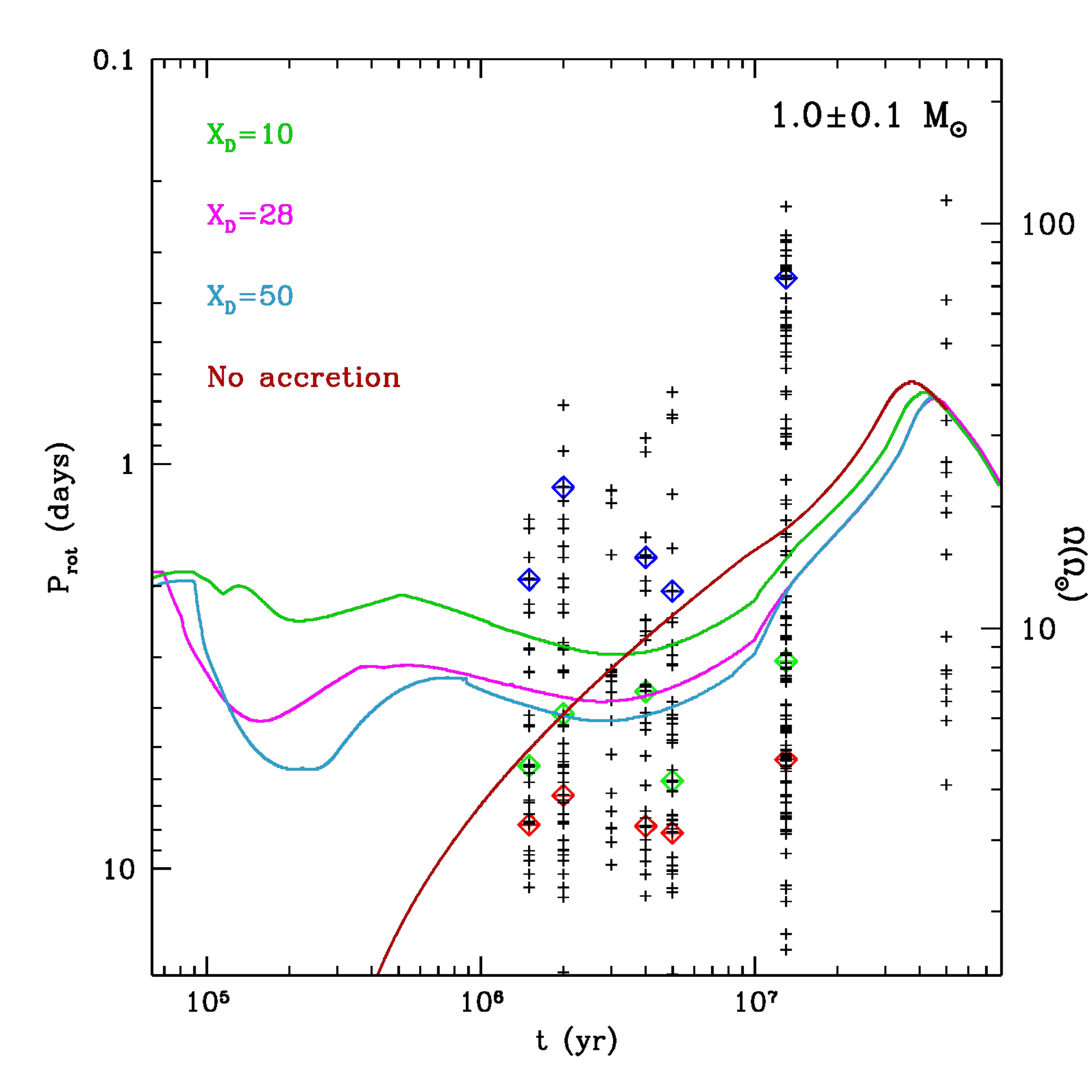}\\
    \includegraphics[width=0.42\textwidth]{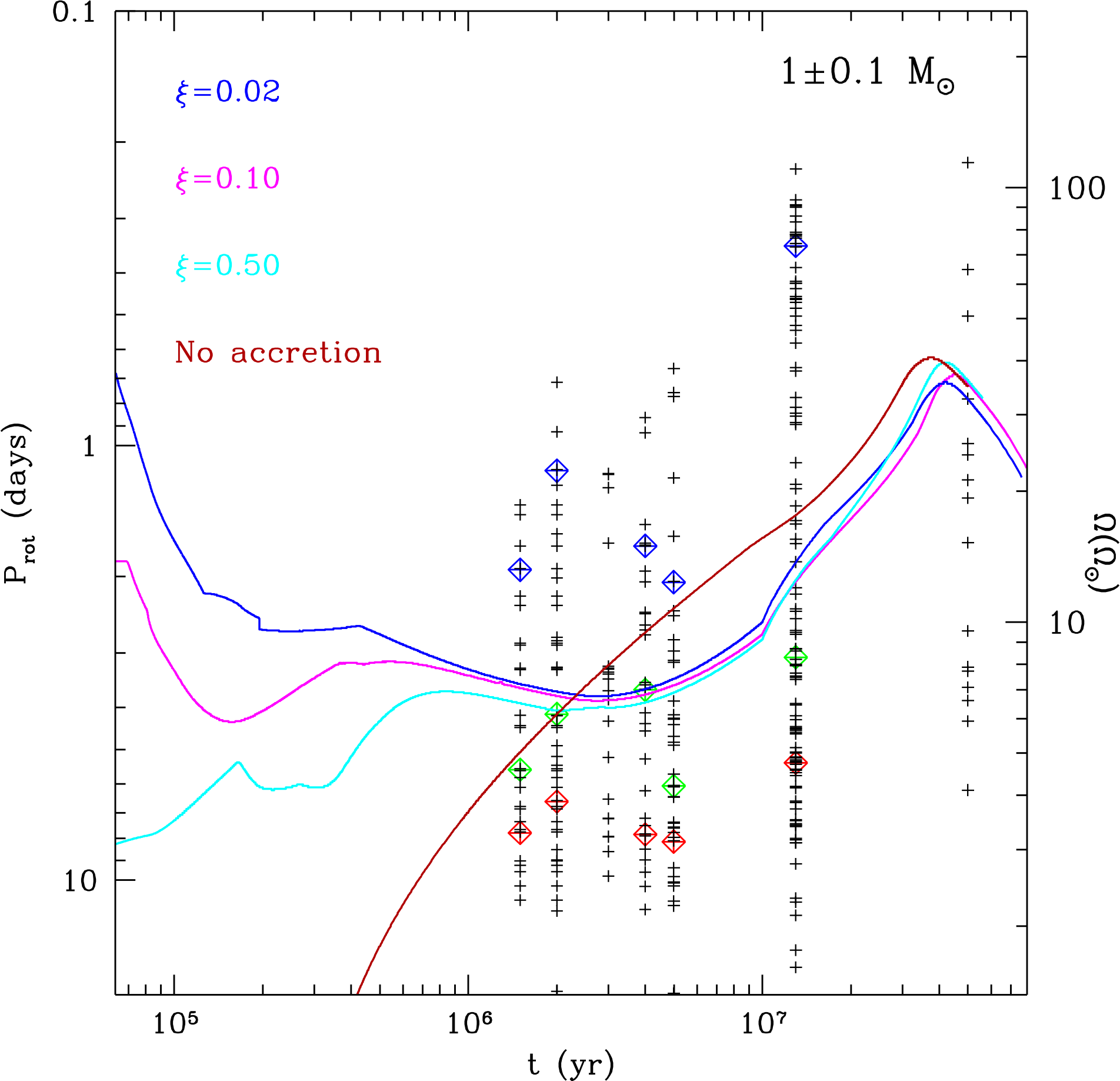}\\
    \includegraphics[width=0.42\textwidth]{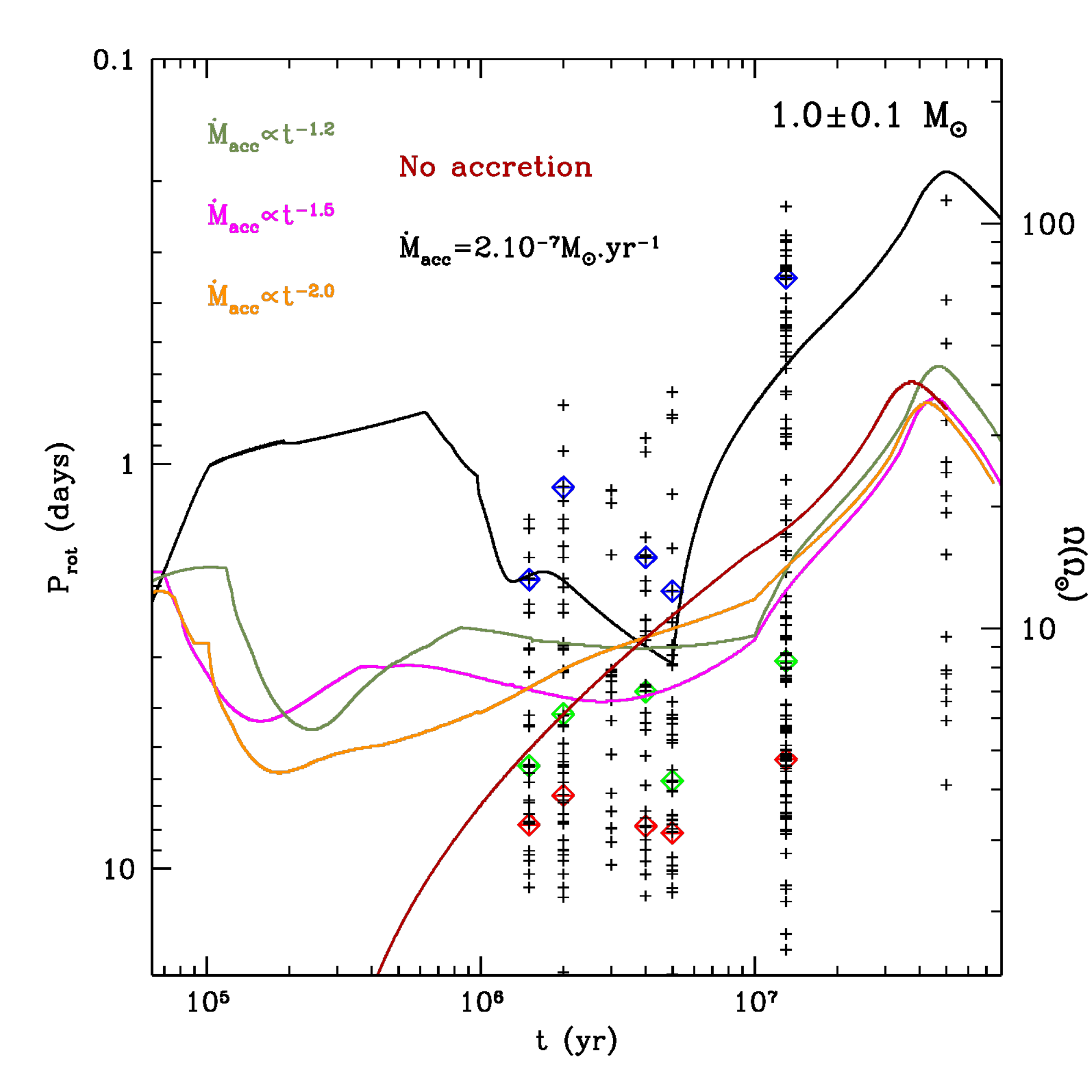}
    \caption{Evolution of the surface rotation period as a function of time when varying the amount of Deuterium in the accreted material (top), the entropy in the accreted material (middle), and the mass accretion history (bottom). The black crosses show observed rotation periods in very young open clusters. The red, green and blue diamonds indicate the 25th percentile, the median and the 90th percentile of each distribution with a sufficient number of stars  (see text for details). }
    \label{fig:obsrot_D_E_Macc}
\end{figure}

\subsection{Exploring accreted-to-ejected mass ratio and dipole strength}

The torques depend on two additional parameters that we vary in this section: the magnetic strength, which we take as fixed value at the stellar pole, and the ratio between the mass-loss rate and the mass-accretion rate. 
In Figure~\ref{fig:obsrot_B}, we present three models with dipole strength between 750 and 2000G comparable to values observed in T-Tauri stars. 
As expected, increasing the dipolar magnetic field strength enhances the spin-down torque associated both with the enhanced winds and the SDI (see Eq.~\ref{Eq:torqueME} and \ref{Eq:torquewinds}). 
This spread alone in magnetic field strength allows to cover about a quarter of the spread in velocity displayed by the observations (on a logarithmic scale). 

Figure~\ref{fig:obsrot_Mloss} shows computations with two different values for the ratio $\dot{M}_{\rm wind}/\dot{M}_{\rm acc}$, but despite a factor 5 between the two cases, both spin evolution remains very close to each other. The difference  only appears toward the end of the disc phase when the wind torque dominates the torque over the SDI (see fig.~\ref{fig:Alltorques}. Thus the case with a higher $\dot{M}_{\rm wind}/\dot{M}_{\rm acc}$ ratio keeps a slower rotation rate because it counteracts more efficiently the effect of the contraction on the spin. 

\begin{figure}
    \centering
    \includegraphics[width=0.42\textwidth]{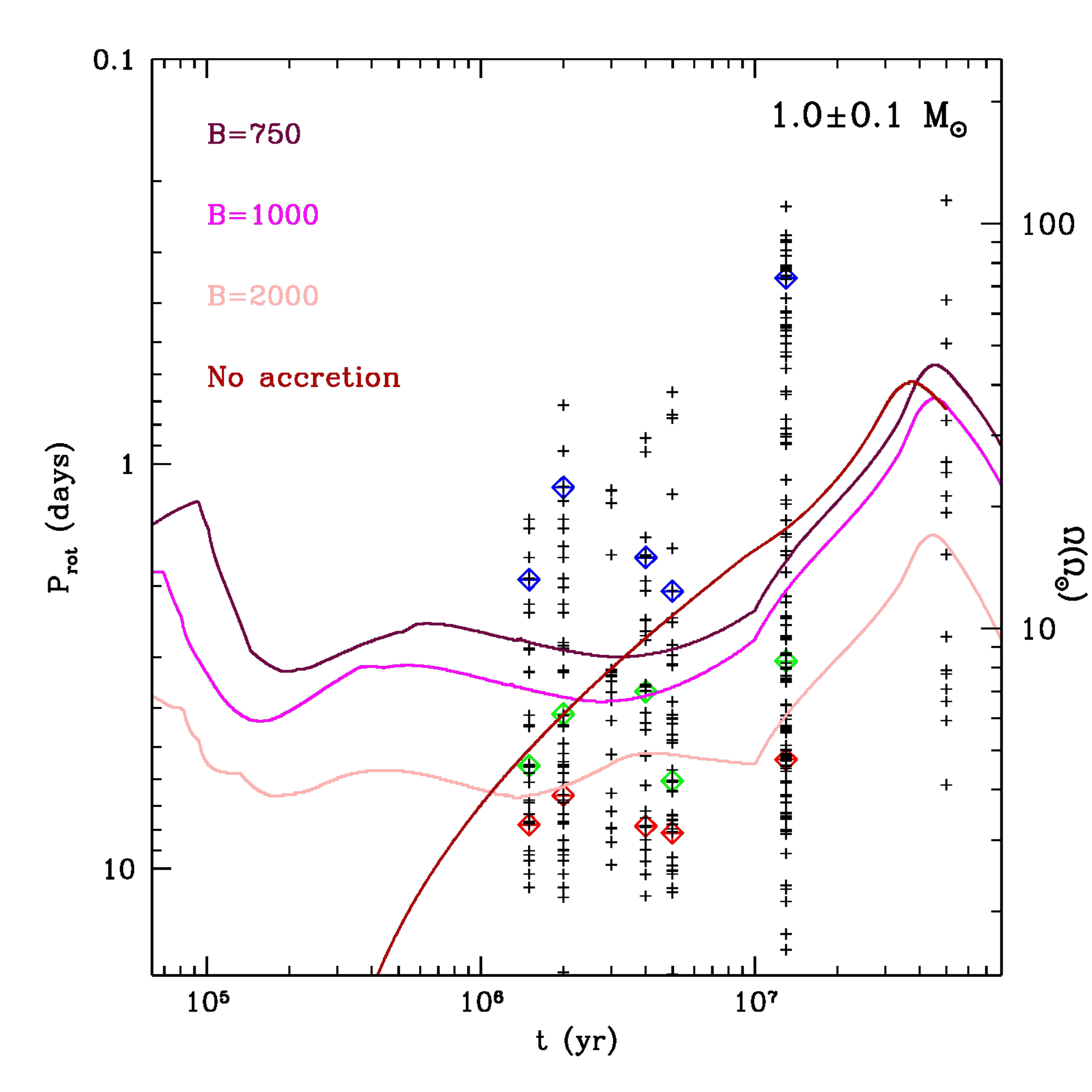}
    \caption{Same as Figure~\ref{fig:obsrot_D_E_Macc} for models with different dipolar magnetic field strength }
    \label{fig:obsrot_B}
\end{figure}

\begin{figure}
    \centering
    \includegraphics[width=0.42\textwidth]{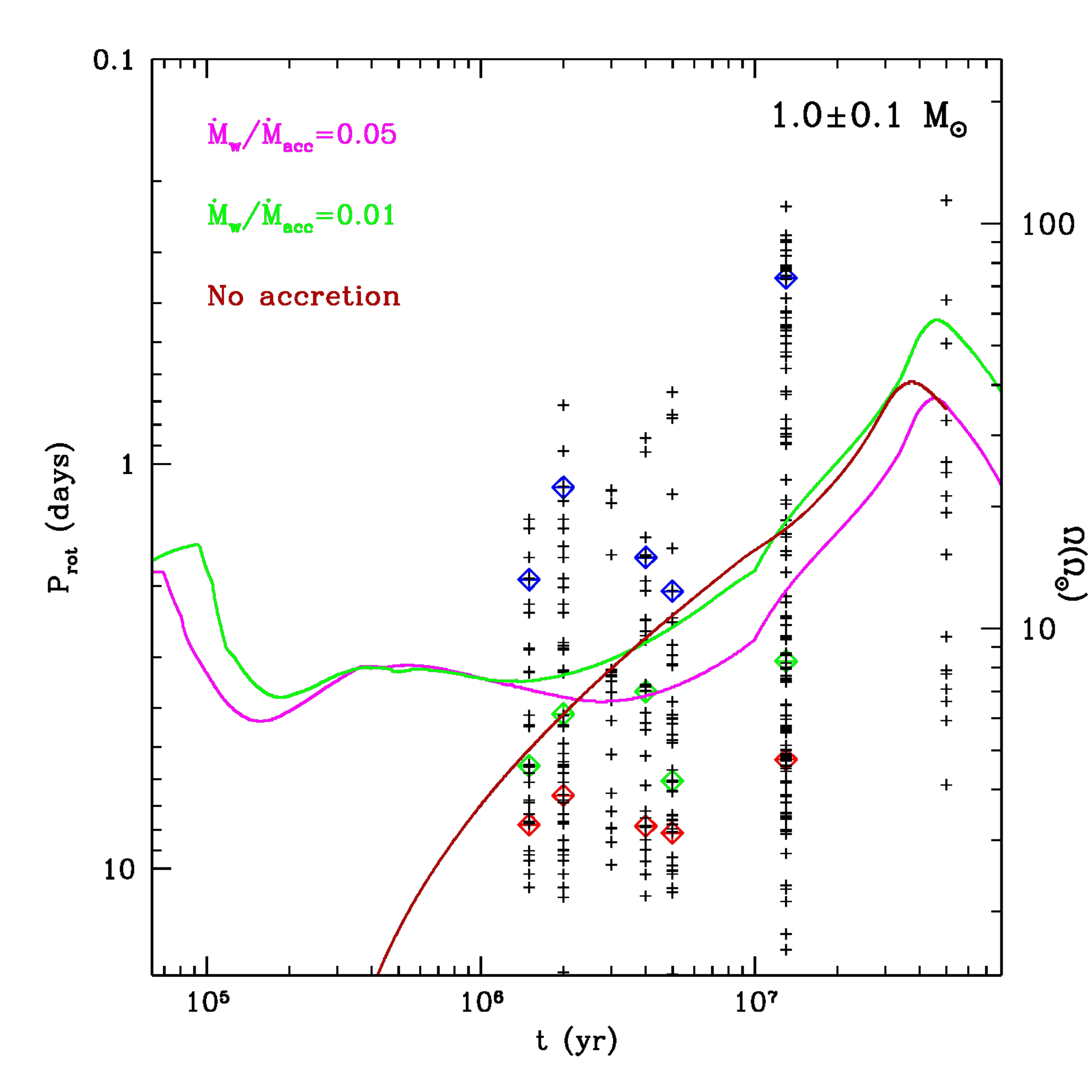}
    \caption{Same as Figure~\ref{fig:obsrot_D_E_Macc} for models with different $\dot{M}_\textrm{w}/\dot{M}_\textrm{acc}$ ratio. }
    \label{fig:obsrot_Mloss}
\end{figure}

\section{Discussion and Conclusion}
Previous stellar evolution models that included the effects of accretion were shown to reproduce the spread observed in colour-magnitude diagrams of young clusters \citep{Hartmann1998,Siess1999,Baraffe2009,Hosokawa2011,Kunitomo2017}. 
In this work, we used the most up-to-date SDI torques from 2.5D MHD simulations, and stellar evolution models that include the effect of both accretion and rotation to model the rotation period distributions of these very young open clusters. 
We show in particular on figures~\ref{fig:obsrot_D_E_Macc}, \ref{fig:obsrot_B}, and \ref{fig:obsrot_Mloss} that their rotational evolution track roughly through the middle of the rotation rate distributions measured in open clusters, during the last few million years of the accretion phase, and evolving in a way that is similar (within the large scatter) to the observed distributions. 
These physics-based models provide a more meaningful interpretation of the observed spin distributions than the large class of spin-evolution models that artificially enforce a constant rotation rate, to avoid the strong spin-up predicted by non-accreting models. 
Our study of this phase revealed that the structural evolution is a critical factor that significantly alters both the moment of inertia and the external torques related to the star-disc interaction. Therefore, in order to model the early spin-evolution, it is necessary to take all of these effects into account, in a self-consistent way.

The changes associated with deuterium burning are particularly important, since the continuous refuelling means that its effects lasts for the whole accretion phase. 
In addition, in agreement with what was found in some early works on stellar formation, the complex processes involving burning of newly added deuterium significantly alters a star's moment of inertia, compared to the expectation from non-accreting models, for example by preventing the star from becoming fully convective.
We find that with a larger deuterium content, the star is spun down more efficiently during the disc coupling phase without affecting the rotation on the ZAMS. In other words, changing the deuterium content mostly changes the rotation rate during the star-disc interaction phase, without changing the final angular momentum content.
Similarly, the accretion rate history plays a critical role in shaping the rotational evolution of the star.  We showed that an initially higher accretion rate brings the star to a slower rotation rate because of the increased inflation of the star. And later, as the accretion rate decays to zero, the steeper the slope of decay, the more the tendency for a star to spin up with time. 

We also showed that changing the heat content of the accreted material strongly affects the early evolution. As previously showed \textit{e.g.} by \cite{Kunitomo2017}, a colder accretion leads to a more compact star early on and therefore, in our case, to a faster rotator. But this difference in evolution does not last after $10^6$ years for the cases we tested. At this point, the thermal energy brought by the accretion becomes negligible, and thus the structure is expected to be independent of differences in the accreted heat content (See fig~\ref{fig:Non_rot_alph}) and driven instead by the burning of deuterium. 
Interestingly, despite being very different at $10^5$ years, by an age of $10^6$ years, the rotation period is driven to an equilibrium by the various torques on a short timescale. This has some important consequence for observations because it suggests that the type of accretion (hot or cold) should not affect the current rotation of observed T-Tauri stars, unless some other parameters comes in the mix (for example, a very low deuterium content may leave a larger role to the heat injection parameter at later time for example).

The variations we explored in the accretion- and SDI-related parameters predicted a spread of modelled rotation rates that covered the central regions of the observed distributions. 
Our design was to explore a relatively extreme range, close to the observationally-constrained boundaries for each parameter.  Between ages of 1.5 and 5 Myr, the predicted rotation range covered by the models goes from about 2 to 8 days. 
In comparison, the rotation rates in the observational datasets we used range from less than a day to 11 days. 
A first possibility to explain this discrepancy would be to hypothesise even more extreme values in the parameter ranges, but more extreme values may be unrealistic. For example, while a 5 kG dipole could easily explain the slowest rotators, such a strong field has never been observed on T-Tauri stars.
Another option is to combine the various parameter ranges, for example having a model with both a lower magnetic field and a lower deuterium should predict a faster rotation rate than the range of models shown. 
Testing this will require computing a large grid of models and require a significant amount of computational resources, such as was done recently by  \cite{Steindl2022,Steindl2022Nature} for the case of non-rotating, accreting stars to explore the seismic signature of accretion.
Finally, we can also consider additional processes that are missing in our model, such as a variation of the duration of the accretion phase \citep[we already know that most stars do not retain a disc for 10Myr from ][]{Roquette2021}, or possible dynamical encounters between stars. 

While these models are an improvement over to previous ones, they still lack some important elements that could play a significant role in the structural and rotational evolution. 
For example, the evolution of accretion does not follow a smooth trend, instead, all dedicated simulations show that the accretion rate can be quite variable with short episodes of accretion hundreds of times more important \citep{Baraffe2009,Hosokawa2011,Audard2014,Vorobyov2015,Vorobyov2017}. 
The coupling of these accretion rate leads to a choppy evolution in the HR diagram, but the global evolution is not too different whether a smooth trend or an episodic accretion is chosen \citep{Baraffe2017,JensenHaugbolle2018,Miley2021,Kunitomo2021,Steiner2021,Steindl2022}. 
Recently, \cite{Gehrig2022,Gehrig2023,Gehrig2023b} have shown that the rotational evolution can also be reproduced in these conditions.
Note that, while the evolution model uses a smooth accretion rate, the SDI torques as well as the wind torque we use, come from a largely episodic and dynamical evolution on short timescales that have been averaged over approximately ten rotation periods \citep{Ireland2021,Ireland2022}.

Modelling the effects of magnetic fields remains a challenge in stellar evolution because it is intrinsically a 3D-problem, and stellar evolution models have to be 1-dimensional to be efficient enough to compute on Gyr timescales. 
The large scale magnetic fields of very young stars may be explained by fast rotation and very large convective cells \citep{EmeriauBrun2017,Zaire2017,Zaire2022}, or they might be fossil remnants from the tendency for magnetic flux conservations during the protostellar cloud collapse \citep[\textit{e.g.}][]{Hennebelle2020}, so even the origin and time-evolution of these fields are poorly understood. 
In addition, only an aligned dipole was considered in \cite{Ireland2022} simulations.
In the case of misalignment between the magnetic field and the rotation axis, simulations of protoplanetary discs show that the inner region of the disc warps and may separate from the rest due to a precessing torque from the star, and lead to a two discs system \citep{Takaishi2020,Romanova21} or other complications. The effects of tilted and more complex stellar magnetic fields might result in significantly different torque scalings from the ones we have adopted here.
Internal magnetic fields can also inhibit the convective motions and have an impact on the stellar structure by expanding the size of the convective cells and thus the stellar radius \citep{Feiden2014,Browning2016,Ireland2018}.
\LA{Finally, the surface magnetic field does not keep a fixed value. Instead, it has been shown to be strongly linked to the stellar Rossby number, which is the ratio of the rotation period and the convective turnover time \citep[e.g.][]{CS2011,Folsom2016,Folsom2018}.}
Most of the models we present possess a radiative core at all times. In these stably stratified regions, the redistribution of angular momentum happens on longer timescale than in the convective zones, and the assumption of instantaneous transport may break down. 
The angular momentum transport timescale depends on the (magneto-)hydrodynamic instability that can develop in the radiative regions, such as hydrodynamic instabilities \citep{Zahn1992,MathisZahn2004,Prat2014}, magnetic instabilities \citep{Bugnet2021,TakahashiLanger2021,Petitdemange2022,Eggenberger2022}, or internal gravity waves \citep{Charbonnel2013,Ahuir2021}, but generally the star ends up in a state of differential rotation.
These instabilities should thus be considered. 

The deposition of mass and energy by the accreted material is likely to vary during the evolution. During the early phase of accretion, material likely falls directly onto the star, rather than being channeled into magnetic accretion columns. \cite{Bandhare2020} show that, during the strong accretion phase, most of the accreted energy is kept by the star, rather than being radiated away, due to the very important opacity of the infalling material, which should lead to a very large stellar radius.
Also, \cite{Siess1997} showed that, in the case of a boundary layer accretion, the mixing in the star depends on the entropy of the accreted material with lower entropy material diving deeper, in agreement with \cite{Geroux2016}'s simulations.

Finally, the chemical composition of the disc evolves spatially and in time as the star evolves \citep[e.g.][]{Miley2021}. When considered, this effect can impact the stellar structure on the main sequence and help to reconcile the helioseismic and spectroscopic properties of the Sun, by changing the composition of the core \citep{Zhang2019,Kunitomo2021}. 
\cite{Gehrig2022,Gehrig2023} showed the importance of the composition when considering rotating stars during the accretion phase and its role on the disc. They find for example that metal-poor stars have shorter disc lifetime with much less efficient torques, letting the star spin-up significantly during the disc coupling phase.

In general, for further progress, future improvements to the models should consider the above effects.

\begin{acknowledgements}
       LA would like to thank Lionel Siess and Isabelle Baraffe for fruitful discussions at the beginning of this work.
       LA acknowledges funding from CNES via an AIM/PLATO grant. LA and SPM acknowledge funding from the European Research Council (ERC) under the European Union's Horizon 2020 research and innovation programme (grant agreement No 682393 AWESoMeStars).  SPM also thanks the University of Washington Astronomy Department for hosting him, as well as the Simons Foundation Flatiron Institute Center for Computational Astrophysics for hosting and support, during part of this work.
\end{acknowledgements}
\bibliography{AccretDL}

\end{document}